\documentclass[preprint,amsmath,amssymb,amsfonts,aps,prab,12pt,obeyspaces,floatfix,superscriptaddress,showkeys]{revtex4-2}
\usepackage{natbib}
\usepackage{graphicx}
\usepackage{lmodern} 
\usepackage{type1cm} 
\usepackage[T1]{fontenc}
\usepackage[normalem]{ulem}
\usepackage{mathtools}
\usepackage{array}
\usepackage{xfrac}
\usepackage{soul}
\usepackage{dcolumn,multirow}
\usepackage[colorlinks=true]{hyperref}
\hypersetup{
    citecolor=green,
    urlcolor=magenta,
    linkcolor=red
}
\usepackage[noabbrev,capitalise]{cleveref}
\usepackage{booktabs,pgfpages,bm,listings,gensymb,latexsym,xcolor}

\usepackage{subcaption}
\usepackage{listings}[language=Python]
\definecolor{codegreen}{rgb}{0,0.6,0}
\definecolor{codegray}{rgb}{0.5,0.5,0.5}
\definecolor{codepurple}{rgb}{0.58,0,0.82}
\definecolor{backcolour}{rgb}{0.95,0.95,0.92}
\lstdefinestyle{mystyle}{
    backgroundcolor=\color{backcolour},
    commentstyle=\color{codegreen},
    keywordstyle=\color{magenta},
    numberstyle=\tiny\color{codegray},
    stringstyle=\color{codepurple},
    basicstyle=\ttfamily\footnotesize,
    breakatwhitespace=false,         
    breaklines=true,                 
    captionpos=b,                    
    keepspaces=true,                 
    numbers=left,                    
    numbersep=5pt,                  
    showspaces=false,                
    showstringspaces=false,
    showtabs=false
}

\usepackage{todonotes}
\usepackage{comment}

\usepackage[section]{placeins}

\renewcommand{\d}{\text{d}}

\newcommand{\pydwa}{\texttt{pyDWA}}

\newcommand{\E}{\mathcal{E}}
\newcommand{\T}{\mathcal{T}}

\newcommand{\optr}{\texttt{TRANSOPTR}}
\newcommand{\invns}{ns\textsuperscript{-1}}

\newcommand{\tprime}{$\dot{\mathcal{T}}$}

\newcommand{\lundjung}{Lund \& Jung \textit{et al.}~\cite{lund-jung_2025}}

\begin{document}

\title{A particle-in-cell model of beam dynamics in a dielectric wall accelerator}

\author{Christopher M. Lund}
\email{christopher.lund@mail.mcgill.ca}
\affiliation{\footnotesize Medical Physics Unit, McGill University Health Centre, 1001~boulevard D\'ecarie, Montr\'{e}al, H4A~3J1, Qu\'{e}bec, Canada}
\affiliation{\footnotesize Department of Physics, McGill University, 3600~rue University, Montr\'{e}al, H3A~2T8, Qu\'{e}bec, Canada}

\author{Paul M. Jung}
\affiliation{\footnotesize TRIUMF, 4004 Wesbrook Mall, Vancouver, V6T~2A3, British Columbia, Canada}
\affiliation{\footnotesize Department of Physics and Astronomy, University of Victoria, Elliott Building, Victoria, V8W~2Y2, British Columbia, Canada}

\author{Jamiel Nasser}
\affiliation{\footnotesize Princess Margaret Cancer Centre, University Health Network, 610~University Avenue, Toronto, M5G~2M9, Ontario, Canada}
\affiliation{\footnotesize Department of Medical Biophysics, University of Toronto, 101~College Street, Toronto, M5G~1L7, Ontario, Canada}

\author{Morgan J. Maher}
\affiliation{\footnotesize Medical Physics Unit, McGill University Health Centre, 1001~boulevard D\'ecarie, Montr\'{e}al, H4A~3J1, Qu\'{e}bec, Canada}
\affiliation{\footnotesize Department of Physics, McGill University, 3600~rue University, Montr\'{e}al, H3A~2T8, Qu\'{e}bec, Canada}

\author{Julien Bancheri}
\affiliation{\footnotesize Medical Physics Unit, McGill University Health Centre, 1001~boulevard D\'ecarie, Montr\'{e}al, H4A~3J1, Qu\'{e}bec, Canada}
\affiliation{\footnotesize Department of Physics, McGill University, 3600~rue University, Montr\'{e}al, H3A~2T8, Qu\'{e}bec, Canada}

\author{Chau Giang Bui}
\affiliation{\footnotesize Princess Margaret Cancer Centre, University Health Network, 610~University Avenue, Toronto, M5G~2M9, Ontario, Canada}
\affiliation{\footnotesize Department of Medical Biophysics, University of Toronto, 101~College Street, Toronto, M5G~1L7, Ontario, Canada}

\author{Thomas Planche}
\affiliation{\footnotesize TRIUMF, 4004 Wesbrook Mall, Vancouver, V6T~2A3, British Columbia, Canada}
\affiliation{\footnotesize Department of Physics and Astronomy, University of Victoria, Elliott Building, Victoria, V8W~2Y2, British Columbia, Canada}

\author{Rick Baartman}
\affiliation{\footnotesize TRIUMF, 4004 Wesbrook Mall, Vancouver, V6T~2A3, British Columbia, Canada}
\affiliation{\footnotesize Department of Physics and Astronomy, University of Victoria, Elliott Building, Victoria, V8W~2Y2, British Columbia, Canada}

\author{Jan Seuntjens}
\affiliation{\footnotesize Princess Margaret Cancer Centre, University Health Network, 610~University Avenue, Toronto, M5G~2M9, Ontario, Canada}
\affiliation{\footnotesize Department of Medical Biophysics, University of Toronto, 101~College Street, Toronto, M5G~1L7, Ontario, Canada}

\begin{abstract}
    Dielectric wall accelerator (DWA) technology has been proposed as a compact, cost-effective alternative to rf accelerators for proton therapy, but its beam dynamics and practical feasibility remain relatively unexplored. In this work, we derive a three-dimensional, time-dependent axisymmetric electromagnetic field model from a prescribed on-wall excitation and implement it as stacked external field elements in the particle-in-cell code Warp. In the absence of experimental DWA beam-transport data, the implementation is cross-checked against a previously developed linear optics model in \optr{} using deliberately idealized beam conditions. Strong agreement is observed between the two models in this regime. The models are then compared for larger transverse and longitudinal emittances, bunch charges up to 1$\times$10$^8e$, and beam parameters representative of a low-energy proton source. The PIC simulations remain consistent with the linear predictions over much of the investigated range, while also identifying wall interactions, longitudinal phase-space distortions, and space-charge induced aberrations. The PIC model represents an intermediate step between linear optics and combined, geometry-specific electromagnetic and particle-transport simulations. It resolves particle-level transport and beam self-fields while retaining an analytical field description that can be varied without committing to a particular DWA structure. More detailed effects may be introduced through externally generated field data or empirical corrections, including models of cell-to-cell coupling, providing a practical basis for increasingly realistic DWA field and beamline studies.
\end{abstract}

\keywords{dielectric wall accelerator; proton therapy; beam optics; particle-in-cell; Warp}

\maketitle

\newpage

\section{Introduction}\label{sec:intro}

Cancer is one of the leading causes of death worldwide~\cite{who2024}. Radiation therapy is indicated for approximately 50\% of cancer patients~\cite{delaney2005-50percent}, and is most commonly delivered using photon beams generated by $<25$~MeV electron linear accelerators (LINACs). In using ionizing radiation to treat cancer, there is an inherent compromise between maximizing tumour dose and minimizing dose to healthy tissues. The unavoidable exposure of healthy tissues carries risks of both acute and long-term side effects, including the development of second cancers. By exploiting the Bragg peak phenomenon~\cite{bragg1904}, proton therapy (PT) enables more conformal dose distributions than conventional photon radiation therapy~\cite{lomax_plancomparison_1999,schulz-ertner_particletherapy_2007,newhauser_secondmalignanices_2011,mohan2022-review,chen2023-proton-comparison,yan_democratise_2023}. This dosimetric advantage has the potential to improve patient quality of life and long-term outcomes.

However, the size and cost of current proton accelerators (generally cyclotron-based) limit access to this modality~\cite{farr_newhorizons_2018, bortfeld_societal_2021}. Coupled with the lack of definitive clinical evidence to support its superiority for many indications~\cite{mohan2022-review,chen2023-proton-comparison}, PT often fails to meet the so-called ``willingness to pay'' threshold~\cite{verma2016-costeffectiveness}. Indeed, only about 1\% of radiation therapy patients actually receive PT~\cite{mohan2022-review}, which in turn limits the accumulation of clinical evidence~\cite{chen2023-proton-comparison}.

\begin{figure}
    \centering
    \includegraphics[width=0.75\linewidth]{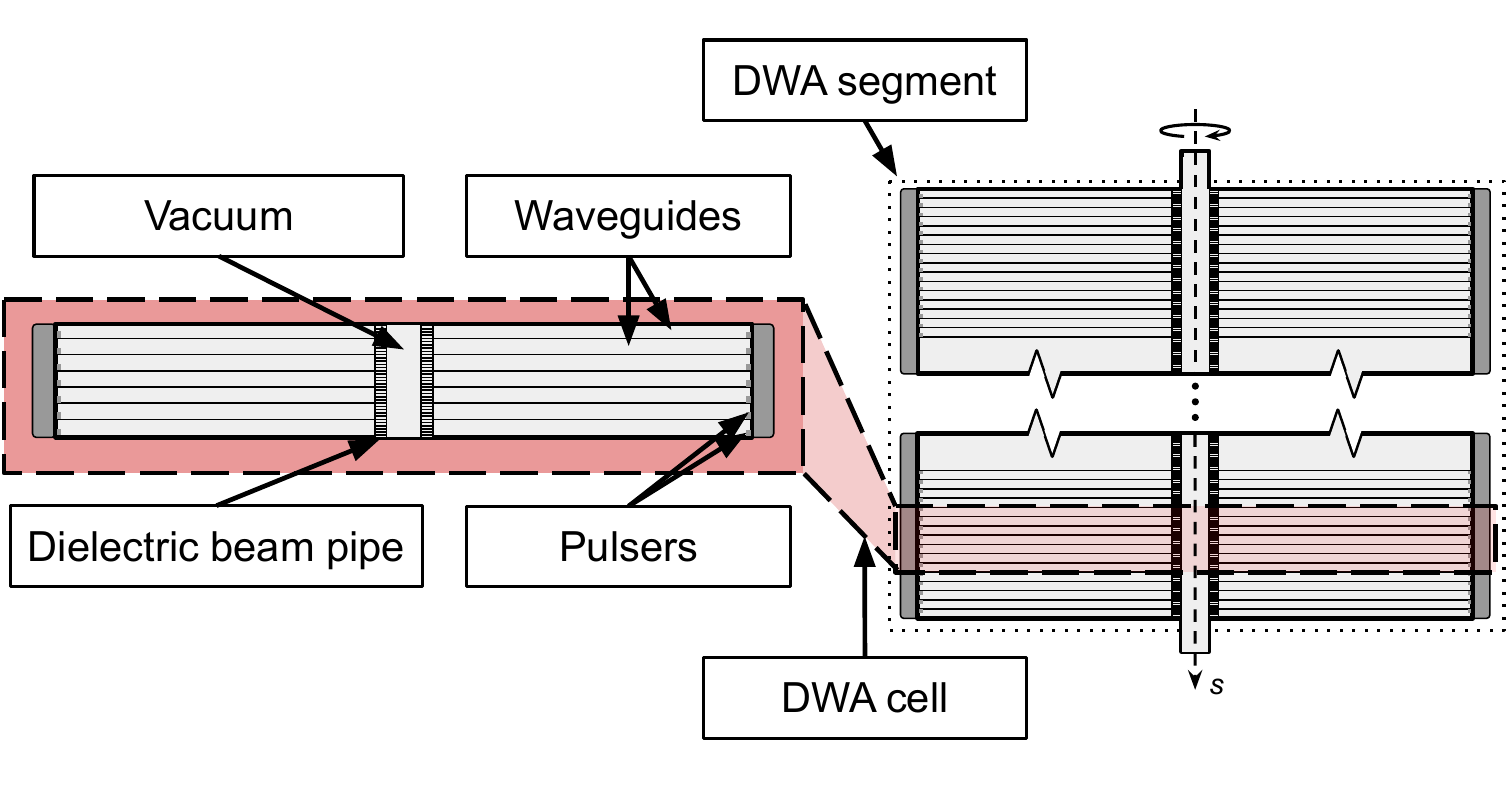}
    \caption{A schematic of a DWA segment. Each cell contains a number of modules, which are all independent field-producing units. Reproduced from \lundjung.}
    \label{fig:dwa-schematic}
\end{figure}

Dielectric wall accelerators (DWAs) are a class of non-resonant linear accelerators that have been proposed for use in proton therapy~\cite{caporaso_compact_2007, caporaso_compact_2008}. These devices are constructed from a large number of independent acceleration modules, each producing a short-lived electric field that extends into the beam pipe. Unlike conventional induction accelerators, which rely on magnetic cores to sustain accelerating fields across discrete gaps, DWAs employ a dielectric beam pipe that permits direct field coupling into the beam region. This enables significantly higher effective accelerating gradients, though at the cost of susceptibility to dielectric breakdown at the dielectric-vacuum interface. If each individual pulse is restricted to nanosecond timescales, this risk can be mitigated, allowing acceleration gradients upwards of 100~MV/m~\cite{nunnally2003}. By coordinating these pulses, particle bunches experience a quasi-continuous acceleration field, an approach referred to as the virtual travelling wave. In principle, this enables DWA systems capable of accelerating protons to clinically relevant energies (up to 250~MeV) within only a few metres. Combined with the absence of large bending magnets, DWA-based systems offer a potential pathway toward compact and cost-effective proton therapy. A basic DWA configuration is depicted in \cref{fig:dwa-schematic}.

At present, DWAs remain largely conceptual. Small-scale prototypes have been demonstrated~\cite{zografos_engineering_2011,zografos_development_2013,shen_psla_2022}, and a number of design challenges have been addressed in recent work~\cite{shen2017,bancheri_2024_dsrd,maher_2025_waveguides}. However, significant work remains before a clinically viable system can be realized. A key step in this process is the development of accurate beam dynamics models. Such models not only inform accelerator design, but also support downstream applications including Monte Carlo dose calculations, treatment planning system development, and shielding analysis. In this way, beam physics simulations provide a critical link between hardware design and clinical implementation.

Moreover, DWAs present several important beam physics challenges. The small beam-pipe aperture, high field gradients, and lack of conventional magnetic focusing create a strong risk of transverse expansion and beam loss. Longitudinally, the flat-top pulses produced by Blumlein-based pulse-forming networks provide no intrinsic focusing, placing strong requirements on bunch length, emittance, and space charge at injection~\cite{zhao_injector_2012,zhu_beam_2014,shen_psla_2022}. Introducing longitudinal focusing through time-varying pulse profiles produces transverse defocusing~\cite{lund-jung_2025}. No established solution therefore exists for maintaining stability in both planes, nor are sufficiently detailed models available to determine which effects will ultimately dominate.

As an initial step towards understanding and designing DWA-based systems, we previously developed a linear optics model~\cite{lund-jung_2025} based on an analytical description of the electromagnetic fields and implemented it in the open-source envelope code \optr{}~\cite{heighway_transoptr_1981}. In the linear optics model, the reference-particle dynamics and second moments of the beam distribution are propagated through linearized fields. Such models enable rapid exploration of accelerator configurations and beam parameters, but they do not resolve the particle distribution directly. As a result, losses, distribution tails, phase-space distortions, and higher-order responses to beam self-fields must either be inferred indirectly or are not represented. Particle-in-cell (PIC) codes can model these effects, albeit at greater computational expense and with increased sensitivity to numerical parameters.

Previous PIC studies of DWAs have generally relied on field maps generated by dedicated electromagnetic field solvers~\cite{zhao_injector_2012,zhu_beam_2014,blackfield_2007_injector,chen_2010_beam}. Such maps are well suited to studying a specific accelerator geometry, but require the structure to be defined in considerable detail. Here, we instead retain the axisymmetric analytical description developed in our previous work. This comes at the cost of neglecting structure-specific asymmetries and other field perturbations, but provides a controlled way to study beam dynamics during prototype development. In doing so, it can highlight effects that should be considered in future designs, guide hardware choices, and provide a testable basis for more detailed simulations before experimental beam-transport data are available.

In this work, we develop a three-dimensional, time-dependent axisymmetric electromagnetic field model for particle-resolved beam transport in DWA structures. Starting from the on-wall excitation used in our previous work~\cite{lund-jung_2025}, we derive the corresponding three-dimensional electric and magnetic fields and implement them as external field elements in the open-source PIC code Warp~\cite{friedman_warp_2014}. In the absence of experimental DWA beam-transport data, the implementation is first cross-checked against \optr{} under deliberately idealized conditions for which the linear and particle-resolved descriptions are expected to closely agree. The model is then applied to selected cases involving beam self-fields, particle loss, and distribution-level departures from linear transport, demonstrating the additional information that PIC simulations can provide for DWA design. The present study considers the beam response to prescribed accelerating fields rather than the complete electrodynamic and pulsed-power behaviour of a DWA system. The particle-transport implementation may also accommodate more detailed simulated or measured field inputs in the future, although obtaining sufficiently complete and validated data will require further development of both DWA technology and its electromagnetic modelling.

\section{Theory}\label{sec:theory}

A DWA segment is composed of a series of independent acceleration modules arranged along the beam axis. These modules may be fired either individually or in grouped sets, referred to as cells. The total field at position $s$ and time $t$ is then the superposition of all active module fields. In our model, each module $i$ is assumed to be identical, such that the field profile need only be derived once and can then be translated in space and time according to the module's midpoint ($s_i,t_i$). For example, the total on-axis field -- which is directed along the beam axis -- can be expressed as~\cite{lund-jung_2025}:
\begin{equation}
    \E_s(s,t) = \sum_{i=1}^{N} \E(s-s_i)\T(t - t_i),
    \label{eq:total_ez}
\end{equation}
where $\E$ is the spatial field profile and $\T$ is a dimensionless temporal profile. While the on-axis description is sufficient for linear optics models, PIC simulations require the fields to be evaluated throughout the beam region. In this section, we derive the time-dependent electric fields in 3D space from the prescribed on-wall excitation and determine the associated magnetic fields.

\subsection{Electric field expansion}\label{ssec:theory-efields}

\noindent To reconstruct the full 3D electric field from the on-axis solution, we seek a multipole expansion of the form:
\begin{align}
\E_r (r,\theta,s) &= - \sum_{\nu=0}^{\infty} \frac{1}{2 \left( \nu +1 \right)} \frac{\d e_{0,\nu}(s)}{\d s} r^{2\nu +1} + \sum_{n=1}^{\infty} \sum_{\nu=0}^{\infty} e_{n,\nu}(s) \left[1+\frac{2\nu}{n}\right] r^{n-1+2\nu} \cos\left[ n \left(\theta + \phi_n \right) \right] \\
\E_\theta (r,\theta,s) &= - \sum_{n=1}^{\infty} \sum_{\nu=0}^{\infty} e_{n,\nu}(s) r^{n-1+2\nu} \sin\left[ n \left(\theta + \phi_n \right) \right] \\
\E_s (r,\theta,s) &= \sum_{\nu=0}^{\infty} e_{0,\nu}(s) r^{2\nu} + \sum_{n=1}^{\infty} \sum_{\nu=0}^{\infty} \frac{1}{n} \frac{\d e_{n,\nu}(s)}{\d s} r^{n+2\nu} \cos\left[ n\left( \theta+\phi_n \right) \right],
\end{align}
where $(r,\theta,s)$ are the normal cylindrical components, $e_{n,\nu}$ is the on-axis field data, $\phi_n$ is phase offset data, and $\nu$ and $n$ are the orders of the expansion in $r$ and $\theta$, respectively. Note that we are only considering the fields as a function of space at this point, and will tack on the time-scale profile when relevant. Assuming axially-symmetric fields, we set both $n=0$ and $\phi_n=0$, and find:
\begin{align}
\E_r &= - \sum_{\nu=0}^{\infty} \frac{1}{2 \left( \nu +1 \right)} \frac{\d e_{0,\nu}(s)}{\d s} r^{2\nu +1} \\
\E_\theta &= 0 \\
\E_s &= \sum_{\nu=0}^{\infty} e_{0,\nu}(s) r^{2\nu}\label{eq:es-expansion}
\end{align}
This assumption removes any geometry-specific asymmetries from the model, including field perturbations from nonuniform pulse propagation in the module waveguides (see \citet{maher_2025_waveguides}). In cases where the fields are expected to be strongly non-axiymmetric in the beam region, additional correction terms or externally supplied field data may need to be considered.

\begin{figure}
    \centering
    \begin{subfigure}{0.49\textwidth}
        \centering
        \includegraphics[width=\textwidth]{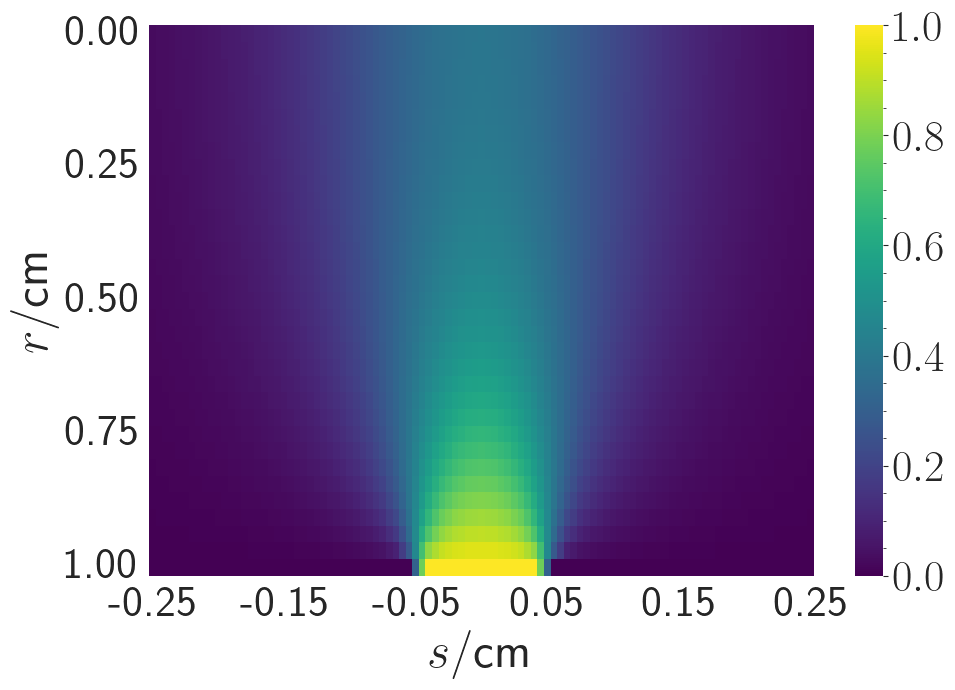}
        \caption{$\mathcal{E}_s/\mathcal{E}_w$}
        \label{fig:efield-calc-ez}
    \end{subfigure}%
    \hspace*{\fill}
    \begin{subfigure}{0.49\textwidth}
        \centering
        \includegraphics[width=\textwidth]{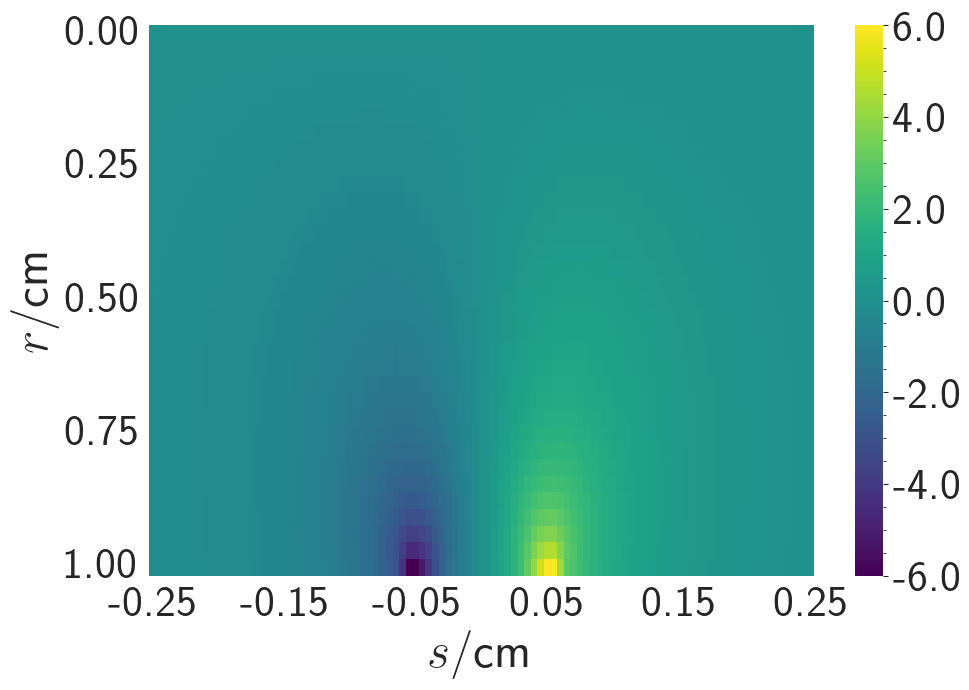}
        \caption{$\mathcal{E}_r/\mathcal{E}_w$}
        \label{fig:efield-calc-er}
    \end{subfigure}%
    \caption{Single module electric field components normalized to the wall field $\mathcal{E}_w$ for a beam pipe radius $r_w = 1$~cm. a) The normalized accelerating field, calculated from \cref{eq:es-field-expansion}. b) The normalized radial field, calculated from \cref{eq:er-field-expansion}. As discussed in the text, the azimuthal field $\mathcal{E}_\theta$ is uniformly 0 and is therefore not shown.}
    \label{fig:efield-calc}
\end{figure}

Our task is now to find an expression for the multipole coefficients $e_{0,\nu}$ and their first-order longitudinal derivatives. From \lundjung, we know that the accelerating field is related to the (known) on-wall excitation field $\E_w$ by:
\begin{equation}\label{eq:es-field}
    \E_s(r,s) = \int_{-\infty}^{\infty} \tilde{\E}_w(k) \frac{I_0(k r)}{I_0(k r_w)} e^{i k s} \d k,
\end{equation}
where \textasciitilde{} represents a Fourier transform, the integral over $k$ is the inverse Fourier transform, $I_0$ is the zeroth order modified Bessel function of the first kind, and $r_w$ is the wall (beam pipe) radius. Taking the Taylor expansion in factors of $r$, we get:
\begin{equation}\label{eq:es-field-expansion}
\E_s(r,s) = \sum_{\nu=0}^\infty c_\nu  \left( \int_{-\infty}^{\infty}\frac{k^{2\nu} \tilde{\E}_w(k)}{I_0(k r_w)}   e^{i k s} \text{d} k \right) r^{2\nu},
\end{equation}
where $c_\nu = (2^\nu \nu!)^{-2}$ are the Bessel function coefficients. An example calculation for $r_w = 1$~cm is plotted in \cref{fig:efield-calc-ez}. Comparing \cref{eq:es-field-expansion} with the general expansion in \cref{eq:es-expansion}, we identify the multipole coefficients as:
\begin{equation}
e_{(0,\nu)}(s) = c_\nu \int_{-\infty}^{\infty}\frac{k^{2\nu} \tilde{\mathcal{E}}_w(k)}{I_0(k r_w)}   e^{i k s} \text{d} k,
\end{equation}
and thus:
\begin{equation}\label{eq:er-field-expansion}
    \E_r (r,s) =  - \sum_{\nu = 0}^{\infty} \frac{c_\nu}{2 (\nu + 1)} \frac{\d}{\d z} \left( \int_{-\infty}^{\infty}\frac{k^{2\nu} \tilde{\mathcal{E}}_w(k)}{I_0(k r_w)}   e^{i k s} \text{d} k \right) r^{2\nu+1}
\end{equation}
An example calculation for $r_w = 1$~cm is plotted in \cref{fig:efield-calc-er}. We can now include the time-scale profile $\T$ and sum over all modules:
\begin{align}
    \E_s (r,\theta,s,t) &= \sum_{i=1}^{N} \E_{s,i}(r,s-s_i) \T(t-t_i)\label{eq:evec-s} \\
    \E_\theta (r,\theta,s,t) &= 0\label{eq:evec-theta} \\
    \E_r (r,\theta,s,t) &= \sum_{i=1}^{N} \E_{r,i}(r,s-s_i) \T(t-t_i)\label{eq:evec-r}
\end{align}

\begin{figure}
    \centering
    \begin{subfigure}{0.49\textwidth}
        \centering
        \includegraphics[width=\textwidth]{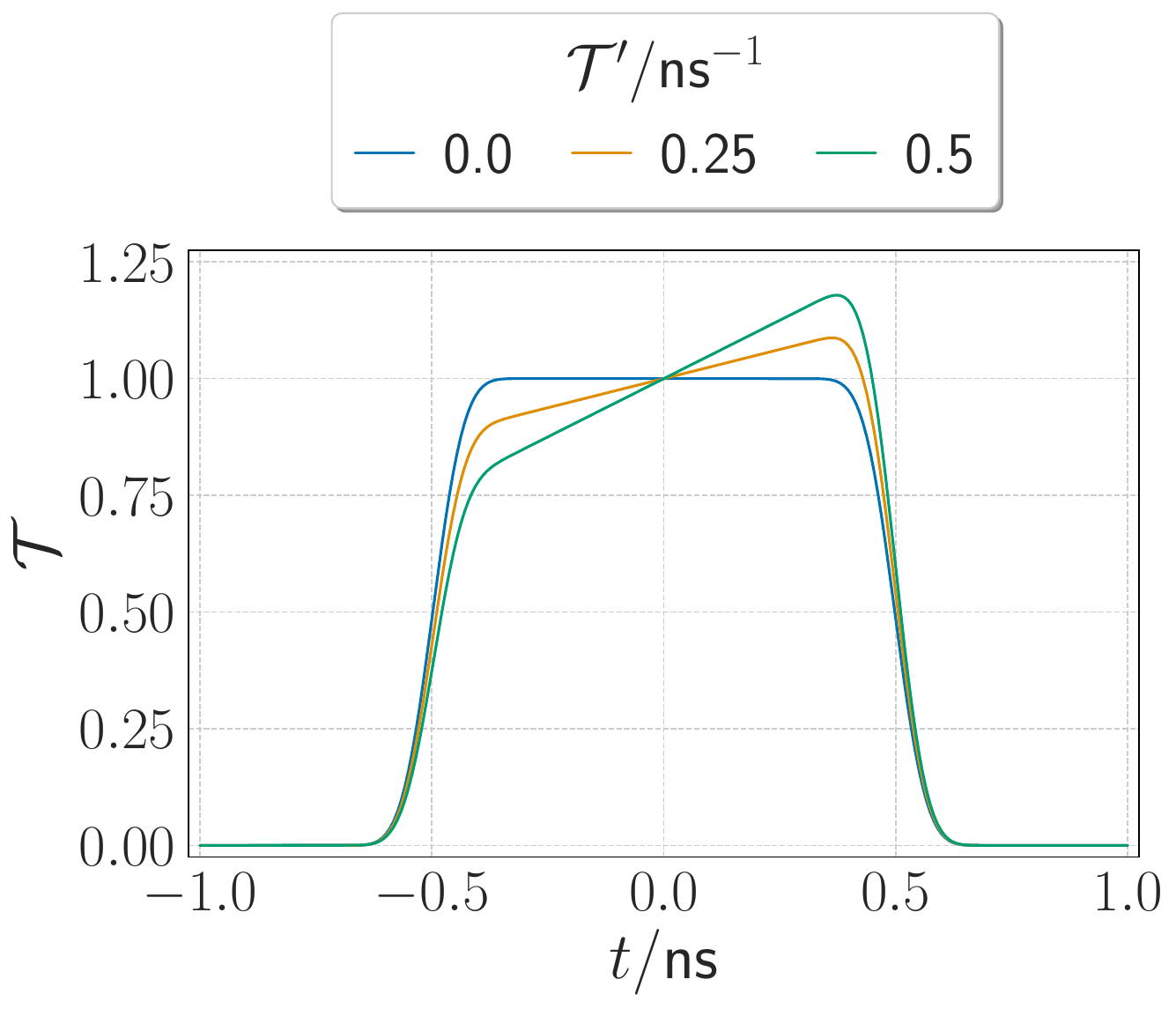}
        \caption{}
        \label{fig:time-scale-profiles}
    \end{subfigure}%
    \hspace*{\fill}
    \begin{subfigure}{0.49\textwidth}
        \centering
        \includegraphics[width=\textwidth]{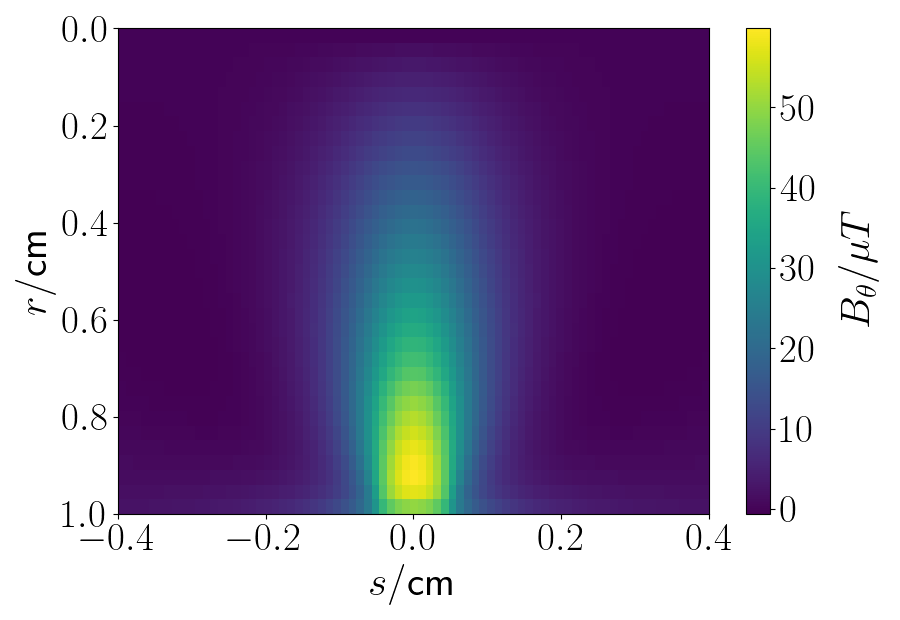}
        \caption{}
        \label{fig:bfield-calc}
    \end{subfigure}%
    \caption{Time variation of the wall excitation and the induced magnetic field. a) Linear time-scale profiles $\T$ as defined by \cref{eq:tscale-profile}. The FWHM was 1~ns and $\dot{\T} = [0, 0.25, 0.5]$~\invns. Reproduced from \lundjung. b) Azimuthal B-field component $B_\theta$ as defined by \cref{eq:bvec-theta}. The wall excitation was $\mathcal{E}_w = 10$~MV/m with $\dot{\T} = 0.25$~\invns{}. The other components are uniformly 0 and are therefore not shown.}
    \label{fig:tscale-btheta}
\end{figure}

Before proceeding, it is worth discussing the form of the temporal profile $\T$ used in this work. As in \lundjung, we will be assuming a pulse whose strength varies linearly with time and whose central value is the nominal voltage:
\begin{equation}\label{eq:tscale-profile}
    \T(t) = 1 + kt,
\end{equation}
where $k \in \mathbb{R}$ is the slope in units of inverse time. Note that this expression is actually a truncation of an expansion of $\T$ about $\Delta t=0$, with our linear assumption setting $\dot{\T} = k$ and $\sfrac{\partial^\textrm{n} \T}{\partial t^\textrm{n}} \approx 0$ for $n>1$. Away from the centre of the pulse, the profile is necessarily non-linear due to finite rise and fall times, and to avoid numerical errors. To accomplish this, the Kaiser smoothing window provided by NumPy~\cite{numpy} was used on the initial hard-edge profile, with the parameters chosen empirically to obtain smooth edges without compromising the linearity of the centre of the pulse. So long as the accelerated bunch is short compared to the pulse length, \cref{eq:tscale-profile} is thus an accurate representation. \cref{fig:time-scale-profiles} shows an example of $\T$ profiles for 1~ns pulses and $k=\dot{\T}=[0,0.25,0.5]$~\invns.

\subsection{Induced magnetic field}\label{ssec:theory-bfields}

\noindent We now account for the magnetic fields induced by the time-varying accelerating field. From Jefimenko's equations, which provide time-dependent generalizations of Coulomb's law and the Biot-Savart law~\cite{griffiths}, the magnetic field may be written as:
\begin{equation}\label{eq:jefimenko}
    \vec{B}(\vec{x},t) = \frac{\mu_0}{4 \pi} \int \left[ \frac{\vec{J}(\vec{x}',t_r)}{|\vec{x}-\vec{x}'|^2} + \frac{\dot{\vec{J}}(\vec{x}',t_r)}{c |\vec{x}-\vec{x}'|} \right] \times \frac{\vec{x}-\vec{x}'}{| \vec{x}-\vec{x}' |} \d \tau',
\end{equation}
where $\vec{x}$ is the position vector in the chosen coordinate space, $\vec{J}$ is the current density, $\d \tau'$ is the infinitesimal volume element, and $t_r$ is the retarded time. Expanding $\vec{J}$ about $t$, truncating to linear order -- valid provided we restrict ourselves to regions away from the pulse edges -- and noting that $t_r = t - \sfrac{|\vec{x}-\vec{x}'|}{c}$ and $\dot{\vec{J}}(t_r) \approx \dot{\vec{J}}(t)$, the bracketed term in \cref{eq:jefimenko} simplifies to:
\begin{equation}
    \frac{\vec{J}(\vec{x}',t_r)}{|\vec{x}-\vec{x}'|^2} + \frac{\dot{\vec{J}}(\vec{x}',t_r)}{c |\vec{x}-\vec{x}'|} = \frac{\vec{J}(\vec{x}',t)}{|\vec{x}-\vec{x}'|^2}
\end{equation}

In the absence of free charges and currents, the effective current density arises from the time variation of the electric field. In vacuum, this may be expressed as the displacement current $\vec{J}=\epsilon_0 \E(\vec{x}) \dot{\T}(t)$. Substitution into \cref{eq:jefimenko} yields:
\begin{align}\label{eq:bfield-crossproduct}
    \vec{B}(\vec{x}, t) &= \frac{\dot{\T}(t)}{4 \pi c^2} \int \d \tau' \E(\vec{x}') \times \frac{\vec{x} -  \vec{x}'}{|\vec{x} -  \vec{x}'|^3} \\
    &= \frac{\dot{\T}(t)}{4 \pi c^2} \int \d \tau' \begin{vmatrix*}
        \hat{x} & \hat{y} & \hat{z} \\
        \E_r (x',y',z') \cos \theta' & \E_r (x',y',z') \sin \theta' & \E_s (x',y',z') \\
        \frac{x-x'}{|\vec{x}-\vec{x}'|^3} & \frac{y-y'}{|\vec{x}-\vec{x}'|^3} & \frac{z-z'}{|\vec{x}-\vec{x}'|^3}
    \end{vmatrix*},
\end{align}
where $\theta' = \arctan(y'/x')$. The cross product is evaluated in Cartesian coordinates for convenience. Exploiting the axial symmetry of the fields, we may set $\theta = 0$ without loss of generality and extend the result to arbitrary $\theta$. Transforming to cylindrical coordinates via $(x,y,z) \to (r,0,s)$ and $(\hat{x},\hat{y},\hat{z}) \to (\hat{r},\hat{\theta},\hat{s})$, the B-field components become:
\begin{align}
B_r(\vec{x}, t) &= \frac{\dot{\T}(t)}{4 \pi c^2} \int \d \tau' \frac{[\E_r(r',\theta',s')\sin\theta'(z-z')-\E_s(r',\theta',s')(-r'\sin\theta')]}{[(r-r'\cos\theta')^2+(-r'\sin\theta')^2+(z-z')^2]^{3/2}} \\
B_\theta (\vec{x}, t) &=\frac{\dot{\T}(t)}{4 \pi c^2} \int \d \tau'  \frac{[\E_s(r',\theta',s')(r-r'\cos\theta')-\E_r(r',\theta',s')\cos\theta'(z-z')]}{[(r-r'\cos\theta')^2+(-r'\sin\theta')^2+(z-z')^2]^{3/2}} \\
B_s(\vec{x}, t) &= \frac{\dot{\T}(t)}{4 \pi c^2} \int \d \tau' \frac{[\E_r(r',\theta',s')\cos\theta'(-r'\sin\theta')-\E_r(r',\theta',s')\sin\theta'(r-r'\cos\theta')]}{[(r-r'\cos\theta')^2+(-r'\sin\theta')^2+(z-z')^2]^{3/2}}
\end{align}
The integrands in the expressions for $B_r$ and $B_s$ are odd functions due to the $\sin \theta'$ terms, and therefore vanish upon integration. The only non-zero component is $B_{\theta}$, consistent with the axial symmetry of the system. The induced B-field -- to a linear approximation in \tprime -- is thus:
\begin{align}
B_r(\vec{x}, t) &= 0 \\
B_\theta (\vec{x}, t) &= \frac{\dot{\T}(t)}{4 \pi c^2} \int r' \d r' \d \theta' \d s'  \frac{[\E_s(r',\theta',s')(r-r'\cos\theta')-\E_r(r',\theta',s')\cos\theta'(z-z')]}{[(r-r'\cos\theta')^2+(-r'\sin\theta')^2+(z-z')^2]^{3/2}} \label{eq:bvec-theta}\\
B_s(\vec{x}, t) &= 0
\end{align}

\subsection{Code release}

The field model presented in this section has been implemented in an open-source Python package, \pydwa{} (\url{https://gitlab.com/dwa-beam-physics/pydwa}). This package also includes a Python implementation of the pulse timing routine described in \lundjung, as well as some classes for data analysis and plotting. The field elements in \pydwa{} are all accessible as spline objects that can be evaluated at specific points or interpolated onto arbitrary grids. 

\section{Methods}\label{sec:methods}

In this work, we assess the analytical field model and its implementation in Warp in two stages. First, we compare the Warp simulations with the previously developed \optr{} model under deliberately idealized conditions where the beam transport is expected to remain predominantly linear. This cross-check tests whether the two implementations reproduce the same transport within this regime. Second, we examine cases with larger transverse and longitudinal emittances, increased bunch charge, as well as a lower-energy beam representative of direct injection from a compact proton source. These simulations are used to demonstrate the additional information available from the particle-resolved model, including particle loss and distortions of the phase-space distributions.

The baseline comparison uses a small, low-emittance bunch with negligible space charge. These conditions are intentionally idealized and are not intended to represent a complete injector design. Rather, they suppress beam-driven effects and interactions with the beam-pipe wall, providing a controlled case in which close agreement between Warp and \optr{} is expected. This set of conditions is referred to as the linear regime throughout this work. The transverse emittance, longitudinal emittance, and bunch charge are then varied independently, allowing the influence of each parameter to be assessed separately.

The DWA segment considered was 15~cm long (in $s$), with a beam pipe radius of 1~cm, and consisted of 150 modules of length 1~mm. Each module produced a 1~ns pulse with a peak field of 10~MV/m. This configuration is a test case representing a short, low-energy acceleration stage and provides a sufficiently simple system for comparison between the two models.

For the baseline cross-check, bunches with an initial energy of 1~MeV were initialized at the entrance to the field region, located 7.5~mm upstream of the first waveguide plate. The normalized 1-rms emittance $\epsilon_n$ was set to $0.016$~$\mu$m in both the transverse ($\epsilon_{n,x}$) and longitudinal ($\epsilon_{n,z}$) planes. The 1-rms bunch radius and temporal length were set to $r = 0.063$~cm and $\Delta t = 0.031$~ns, respectively, corresponding to initial envelopes (2-rms) equal to one-eighth of the beam pipe radius and pulse half-length. These parameters were selected to limit wall interactions and variation across the accelerating pulse, rather than to represent a particular injector. The beam divergence and energy spread were determined from the emittance under the assumptions of an uncoupled beam at waist and in the absence of external fields. The bunch charge was initially set to $q_b = 1e$, effectively neglecting space charge. Simulations were performed for \tprime $= [0.0, 0.25, 0.5]$~\invns{}. Each configuration was simulated using both the \optr{} model and the PIC implementation in Warp. For statistical robustness against finite macroparticle sampling of the initial particle distribution, each Warp simulation was repeated five times with a maximum of 5000 macroparticles per run.

The baseline simulations were then repeated while independently increasing the transverse emittance, longitudinal emittance, and bunch charge. The normalized 1-rms transverse and longitudinal emittances were each increased to 0.25~$\mu$m to examine the effects of increasing transverse beam size and bunch length, including wall interactions and departures from the linear phase-space distributions. Bunch charges up to $1\times10^8e$ were considered to assess the influence of beam self-fields relative to the time-varying DWA fields. When including space charge, the effective envelope corresponds to the $\sqrt{5}$-rms width in the bunched case rather than the 2-rms width~\cite{sacherer1971rms,baartman_2021_bunchedbeams}; the \optr{} simulations were adjusted accordingly.

To provide a less idealized comparison, the same DWA segment was also simulated using beam parameters taken from a TRIUMF-licensed D-PACE proton source~\cite{DPACE}. It was found that transit time effects were leading to significant beam loss, so the DWA pulse duration was subsequently increased to 2~ns. Simulations used 10,000 macroparticles and a source current of 3 mA, with an initial mean kinetic energy of 30~-keV, 1-rms transverse size of $r = 0.05$~cm, a normalized 1-rms transverse emittance of 0.0775~$\mu$m, and a 1-rms energy spread of 0.75\%. The temporal field gradient was set to \tprime{}= 0.25 \invns{}. A 1-rms bunch length of $\Delta t = 0.125$~ns was chosen ($q_b \approx 4.6 \times 10^6 \, e$), but no bunching system was simulated. Note that this test is not intended to represent a complete injector and matching system.

Regarding the PIC simulations, 1590 divisions were used along the beam axis for 150 modules, with each module field element discretized over approximately 150 grid points. Simulations were performed on a cluster running Ubuntu 22.04.2 (LTS) and required on the order of 5 hours on 68-thread compute nodes. Further refinement of the grid produced negligible changes in the results, indicating that the dominant computational cost arises not from resolution, but from the large number of overlapping field elements. This also introduces memory management challenges, necessitating the removal of expired field elements during runtime. An example Warp script is contained in the pyDWA code base.

\section{Results}\label{sec:results}

\begin{figure}
    \centering
    \begin{subfigure}{0.9\textwidth}
        \centering
        \includegraphics[width=\textwidth]{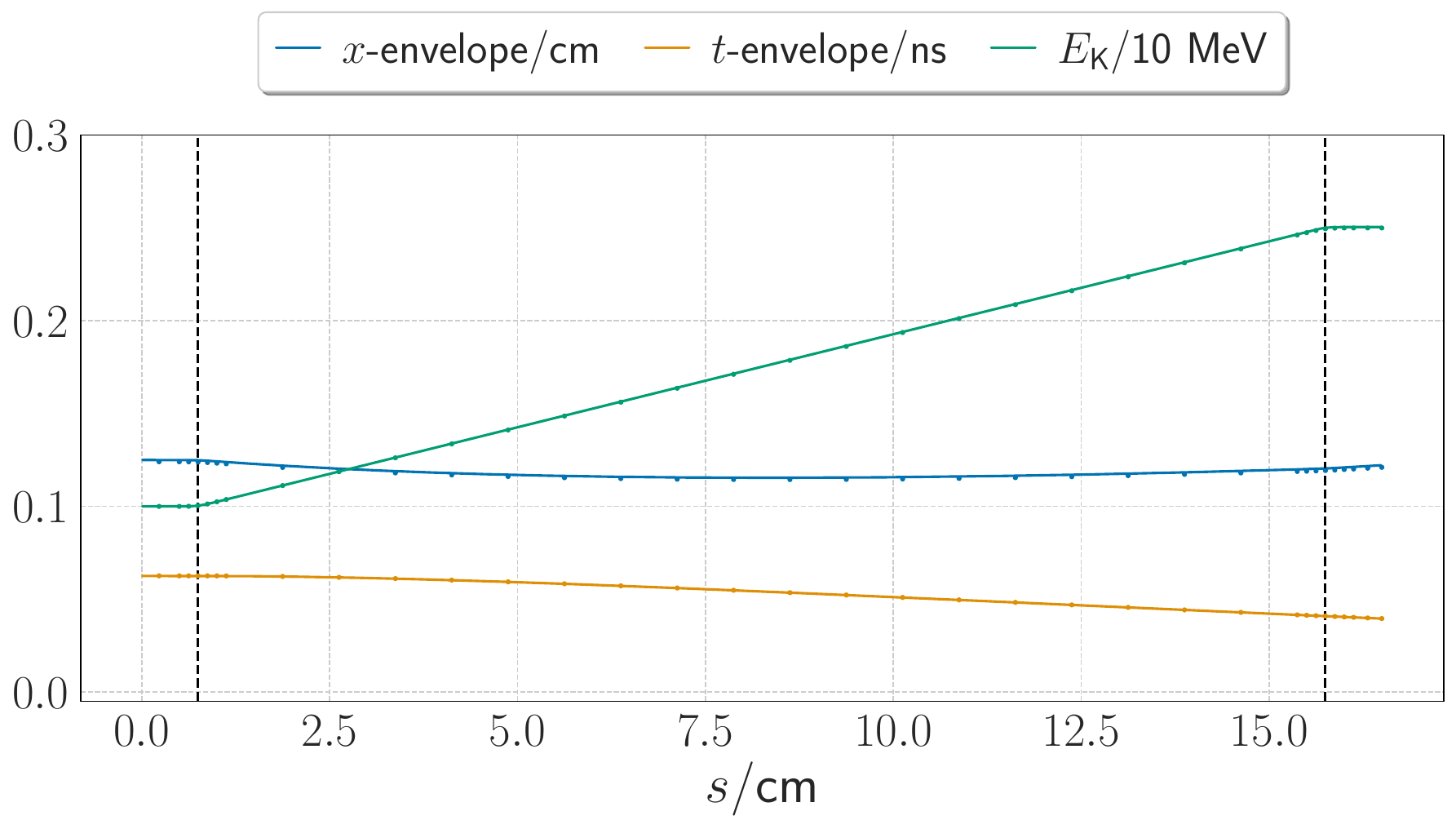}
        \caption{}
        \label{fig:linear-comparison-a}
    \end{subfigure}
    \vspace*{\fill}
    \begin{subfigure}{0.49\textwidth}
     \centering
     \includegraphics[width=\textwidth]{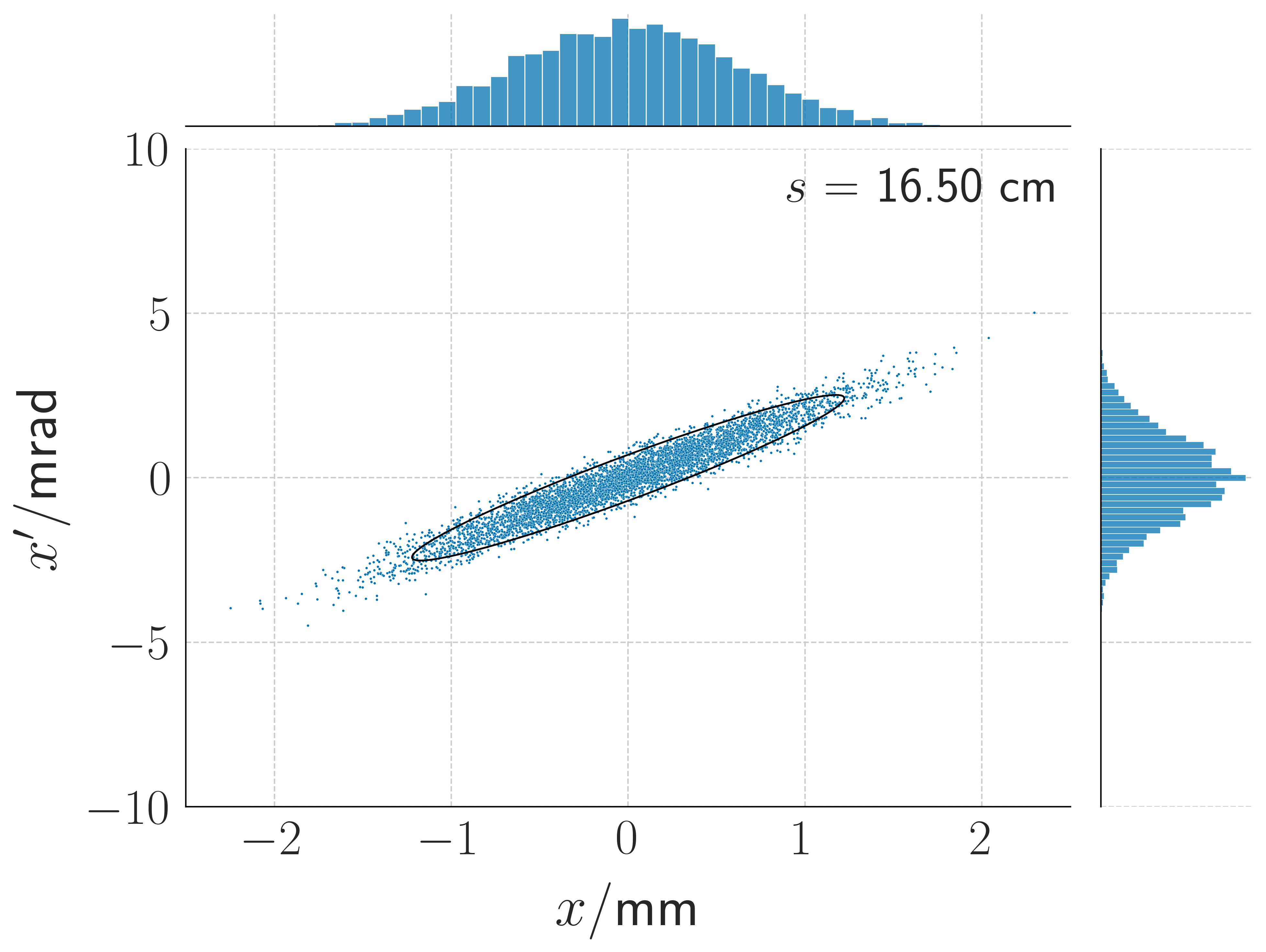}
     \caption{}
     \label{fig:linear-comparison-b}
    \end{subfigure}%
    \hspace*{\fill}
    \begin{subfigure}{0.49\textwidth}
     \centering
     \includegraphics[width=\textwidth]{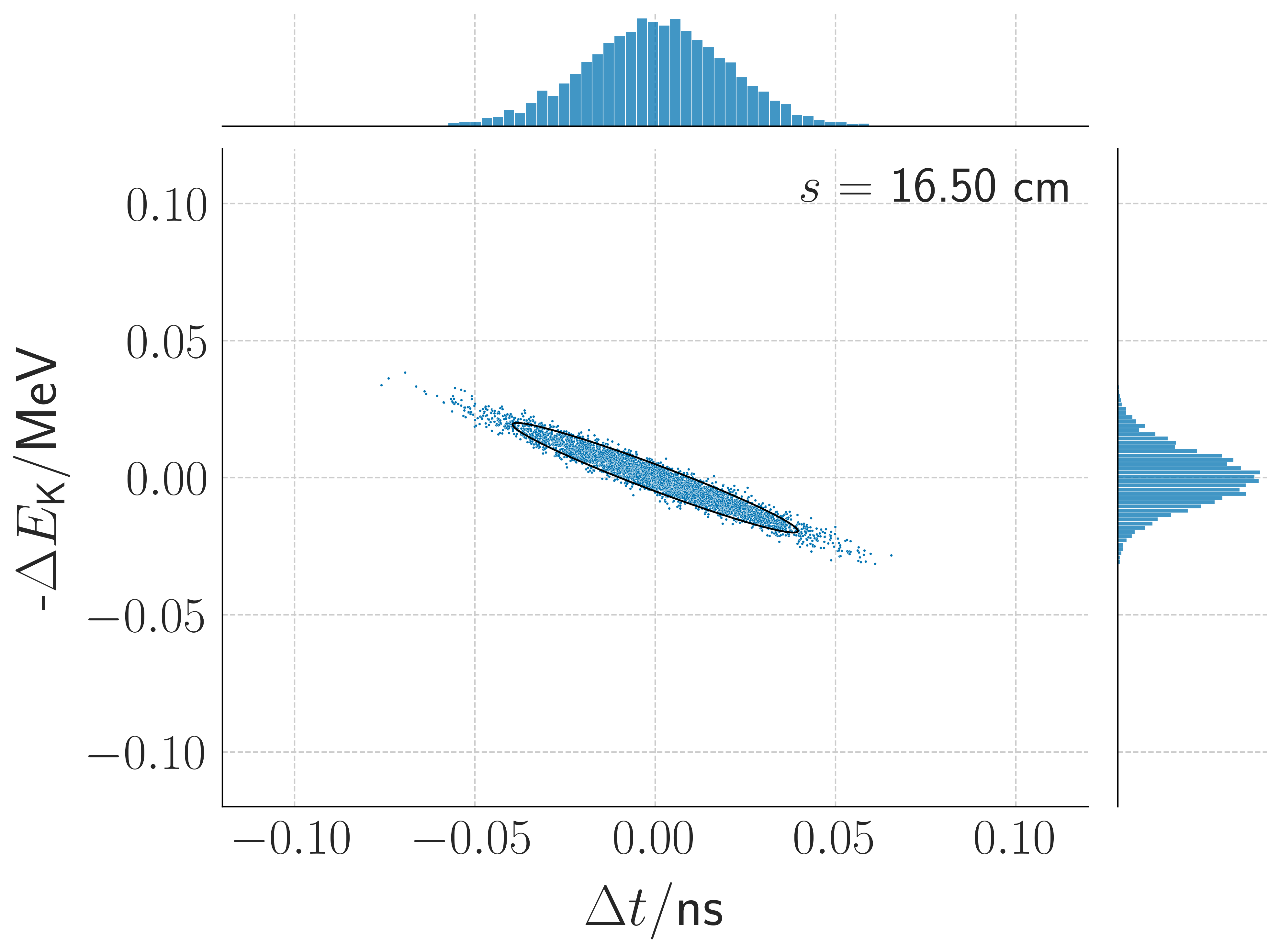}
     \caption{}
     \label{fig:linear-comparison-c}
    \end{subfigure}%
    \caption{Comparison of \optr{} and Warp models of DWA dynamics in the linear regime for a 15~cm, 10~MV/m DWA segment. a) Envelope evolution. Envelopes are 2-rms data from \optr{} (continuous) and Warp distributions (discrete). Warp data is the mean and standard error of the mean over 5 runs, calculated from the particle distributions. b) Transverse phase space at exit. c) Longitudinal phase space at exit. For b) and c), particle data (blue) is from Warp and bounding ellipses (black) are reconstructed from \optr{} envelope data. Histograms above and to the right of each plot show the particle distributions along that phase space axis.}
    \label{fig:linear-comparison}
\end{figure}

\noindent \cref{fig:linear-comparison} compares the Warp and \optr{} models in the linear regime for a 15~cm, 10~MV/m DWA segment with \tprime=0.25~\invns{}, which was taken to be a representative value. In \cref{fig:linear-comparison-a}, the \optr{} envelopes are shown directly, while the Warp envelopes are reconstructed from particle distributions recorded at discrete targets along $s$. The transverse and longitudinal envelopes correspond to the diagonal elements of the 2-rms covariance matrices, and the kinetic energy is given by the mean particle energy at each target. Warp results are reported as the mean and standard error over 5 runs.

In phase space (\cref{fig:linear-comparison-b,fig:linear-comparison-c}), the \optr{} ellipses -- constructed from the corresponding $2\times2$ block diagonal of the beam matrix -- are overlaid on the Warp particle distributions, following the approach of \citet{shelbaya_fast_2019,shelbaya_autofocusing_2021}. Individual macroparticles are shown along with marginal histograms to indicate density.

Excellent agreement is observed between the two models in the linear regime. Transverse and longitudinal envelope deviations are below $\sim 0.9$\% and $\sim 0.1$\% across all $s$, respectively, with kinetic energy agreement at the level of $\sim 0.2$\%. The phase space distributions are consistent in both extent and orientation, and no evidence of higher-order effects is observed in the Warp simulations. These results indicate that the PIC model reproduces the expected linear beam dynamics under conditions where nonlinear effects are negligible.

\begin{figure}[ht]
    \centering
    \begin{subfigure}{0.49\textwidth}
     \centering
     \includegraphics[width=\textwidth]{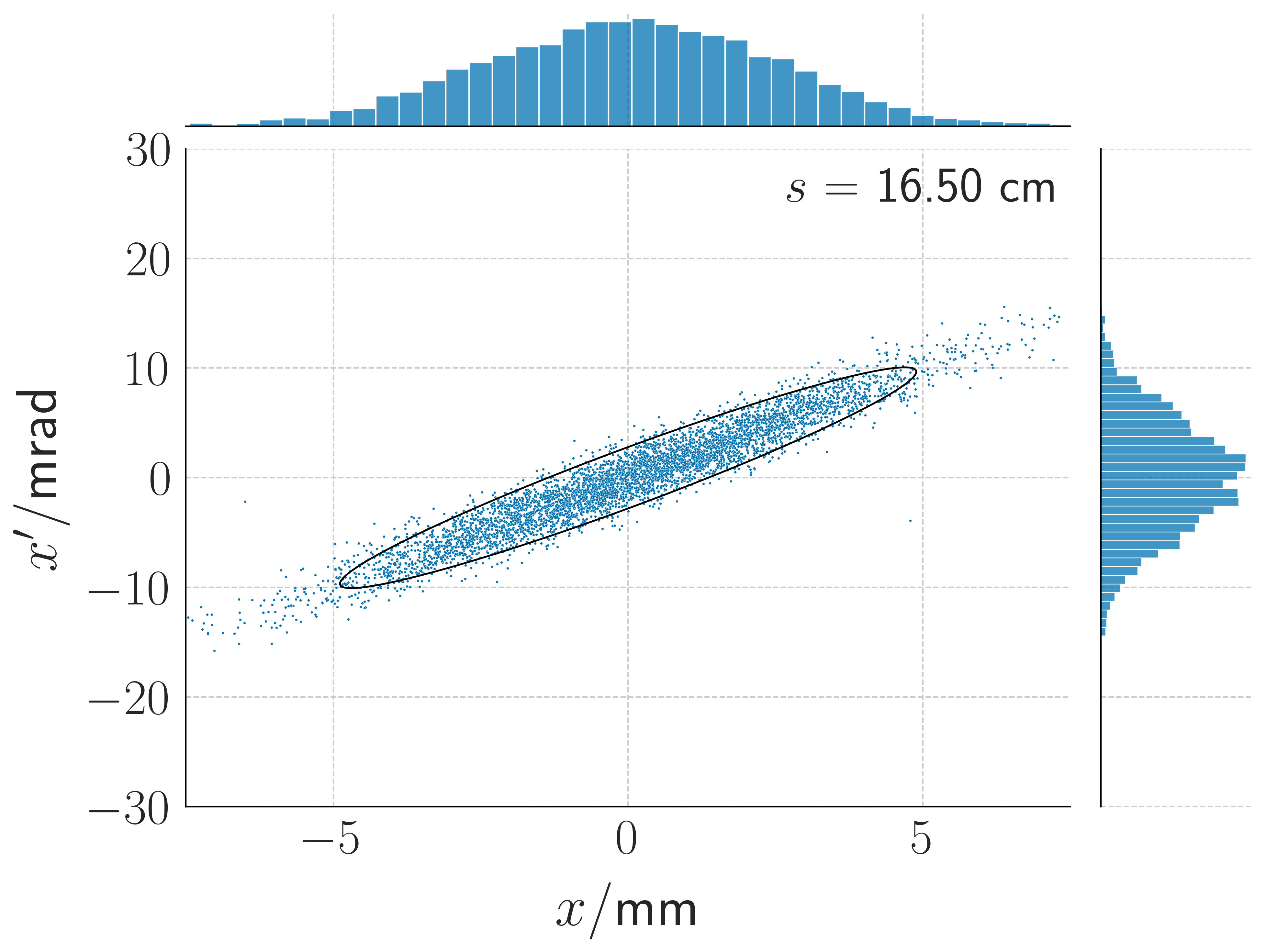}
     \caption{}
     \label{fig:ps-comparison-transverse-wide}
    \end{subfigure}%
    \hspace*{\fill}
    \begin{subfigure}{0.49\textwidth}
     \centering
     \includegraphics[width=\textwidth]{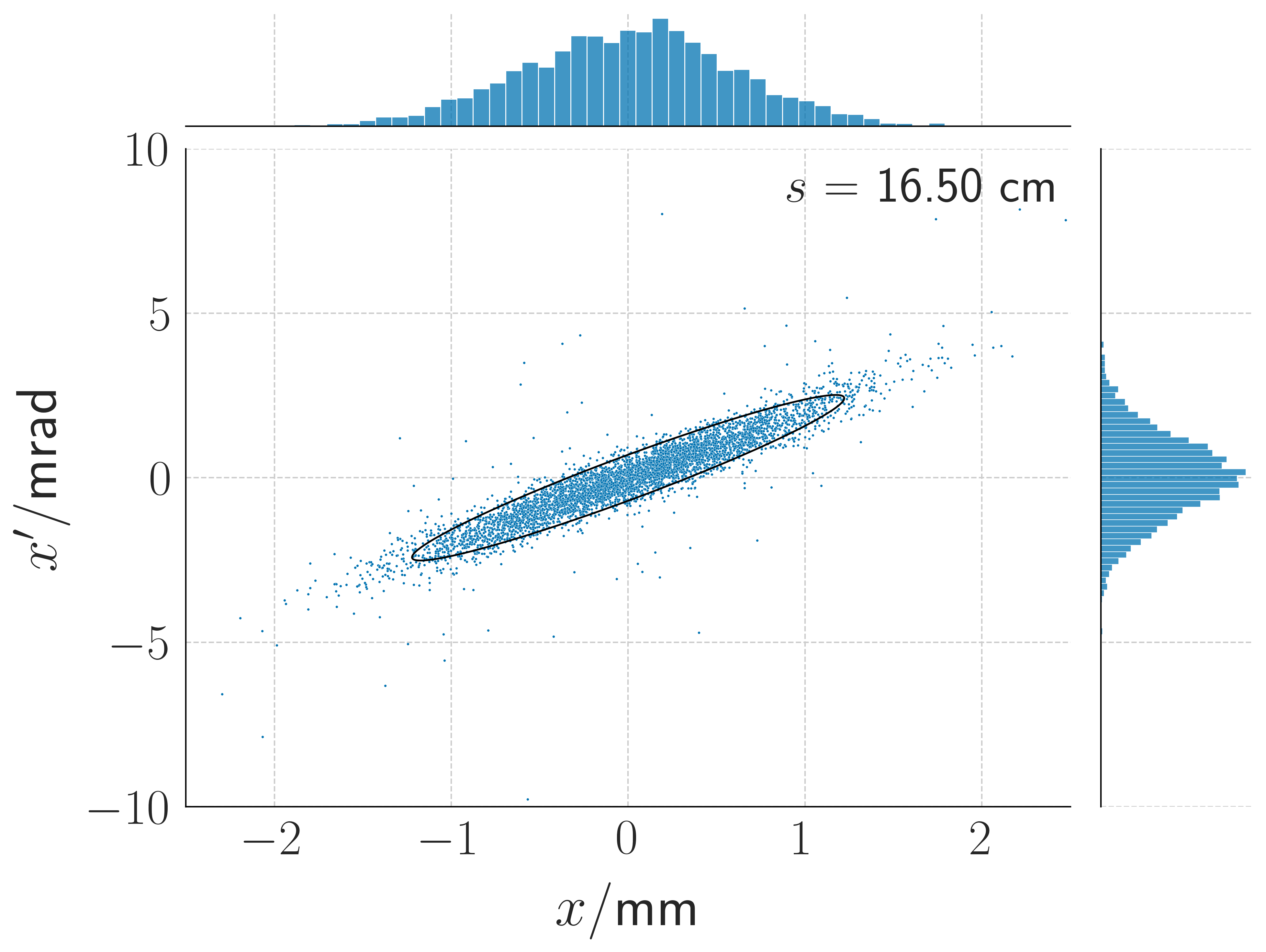}
     \caption{}
     \label{fig:ps-comparison-transverse-long}
    \end{subfigure}%
     \vspace*{\fill}
    \begin{subfigure}{0.49\textwidth}
     \centering
     \includegraphics[width=\textwidth]{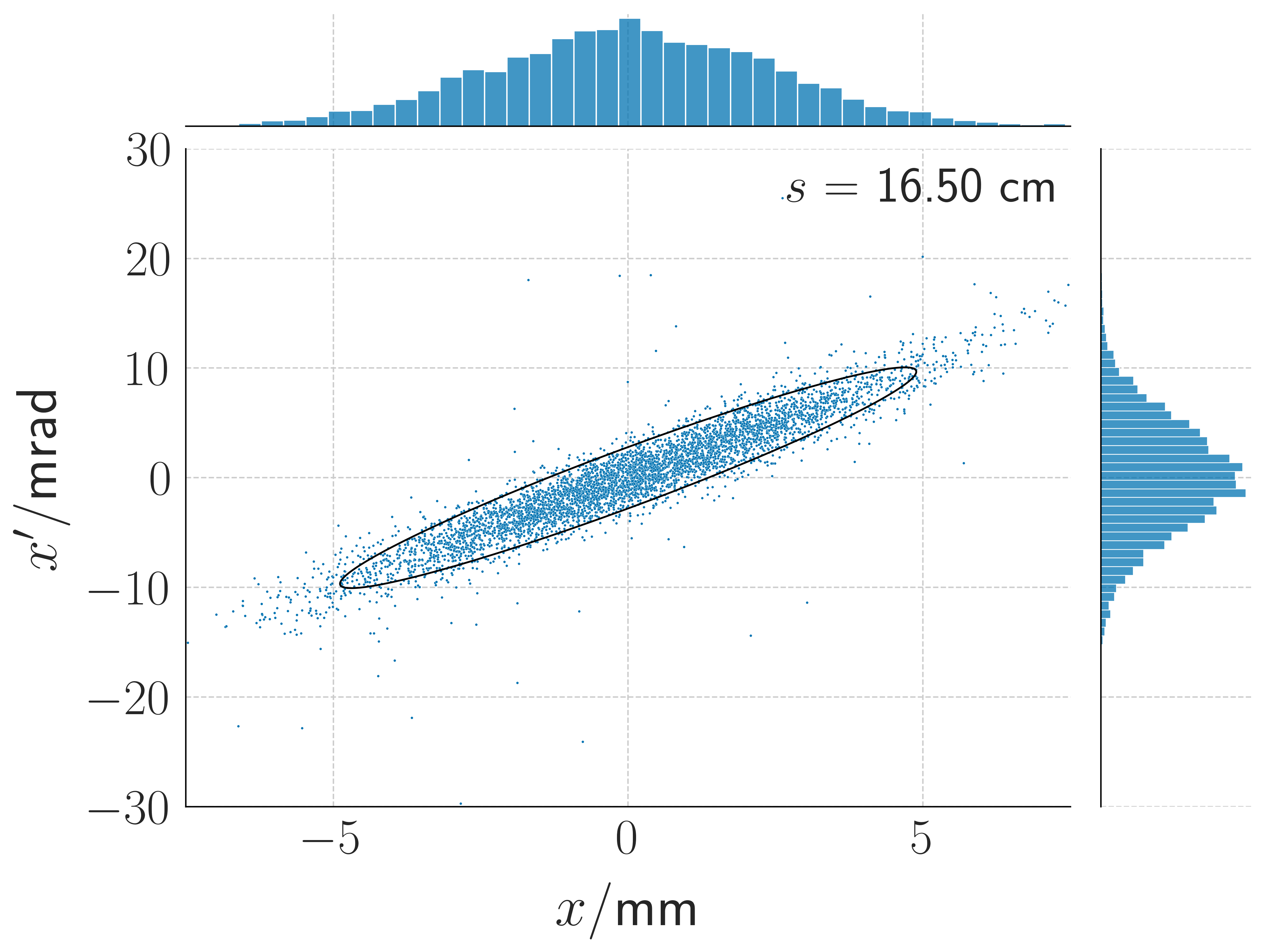}
     \caption{}
     \label{fig:ps-comparison-transverse-1um}
    \end{subfigure}%
    \hspace*{\fill}
    \begin{subfigure}{0.49\textwidth}
     \centering
     \includegraphics[width=\textwidth]{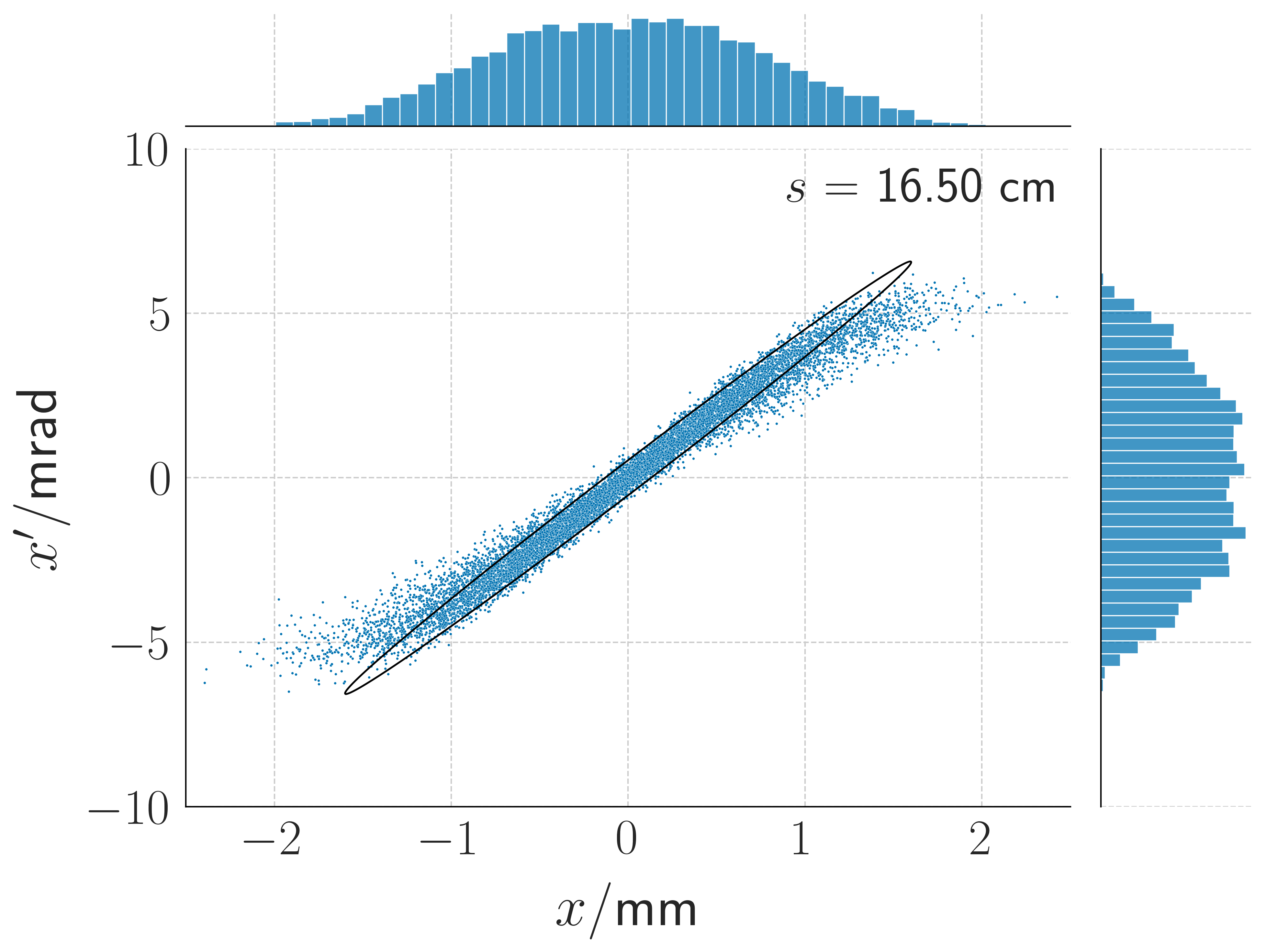}
     \caption{}
     \label{fig:ps-comparison-transverse-sc}
    \end{subfigure}%
    \caption{Transverse phase space at the exit of a 15~cm, 10~MV/m DWA segment. Particle data (blue) is from Warp and bounding ellipses (black) are reconstructed from \optr{} envelope data. Histograms above and to the right of each plot show the particle distributions in $x$ and $x'$, respectively. Each subplot shows the impact of altering a subset of parameters from the baseline rms normalized emittances of $\epsilon_{n,x} = \epsilon_{n,z} = 0.016$~$\mu$m and bunch charge $q_b = 1e$, while keeping the Courant-Snyder parameters constant. a) $\epsilon_{n,x} = 0.25$~$\mu$m, b) $\epsilon_{n,z} = 0.25$~$\mu$m, c) $\epsilon_{n,x} = \epsilon_{n,z} = 0.25$~$\mu$m, d) $q_b = 1\times10^8 e$.}
    \label{fig:ps-comparison-transverse}
\end{figure}

\begin{figure}[ht]
    \centering
    \begin{subfigure}{0.49\textwidth}
     \centering
     \includegraphics[width=\textwidth]{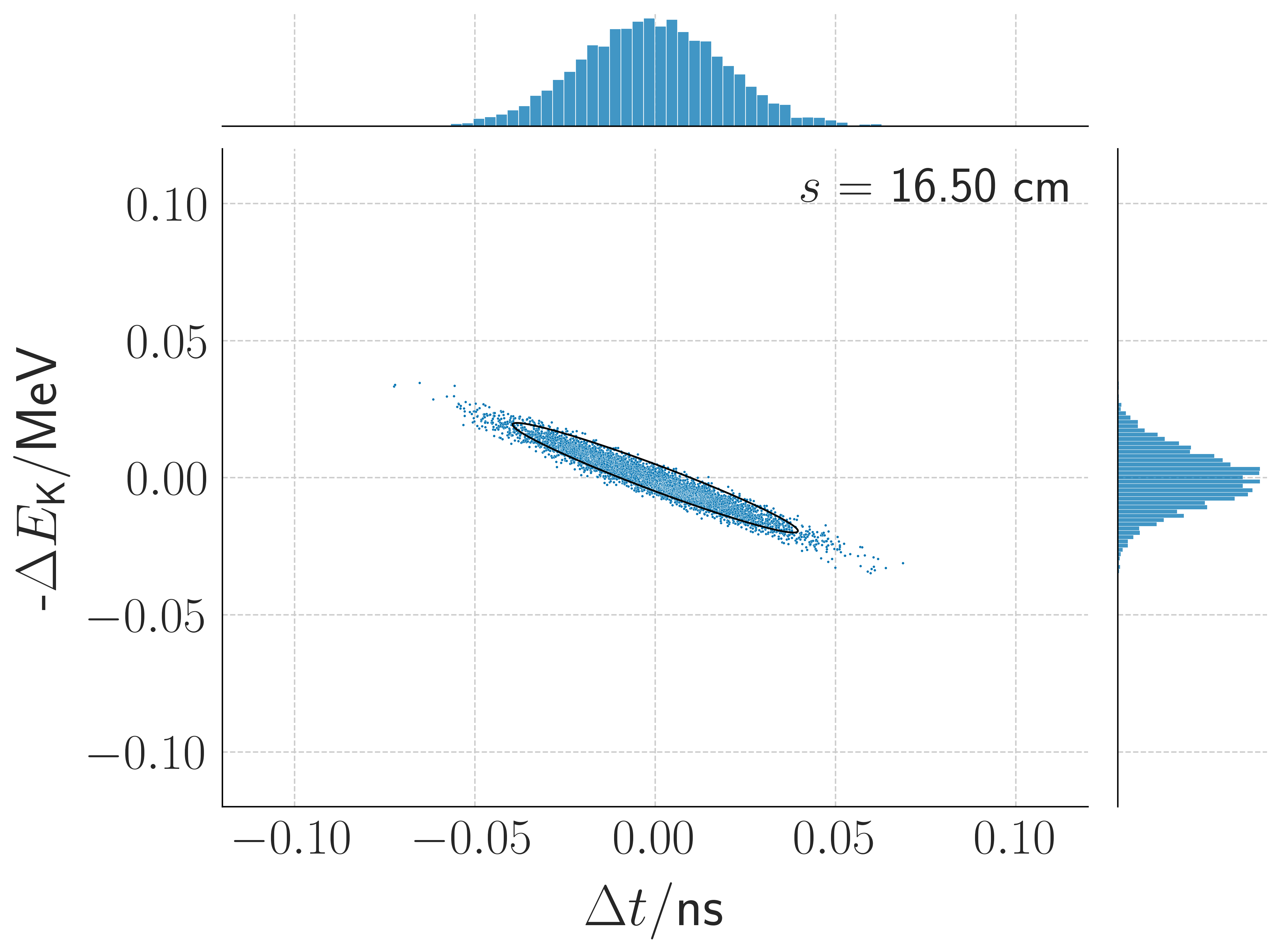}
     \caption{}
     \label{fig:ps-comparison-longitudinal-wide}
    \end{subfigure}%
    \hspace*{\fill}
    \begin{subfigure}{0.49\textwidth}
     \centering
     \includegraphics[width=\textwidth]{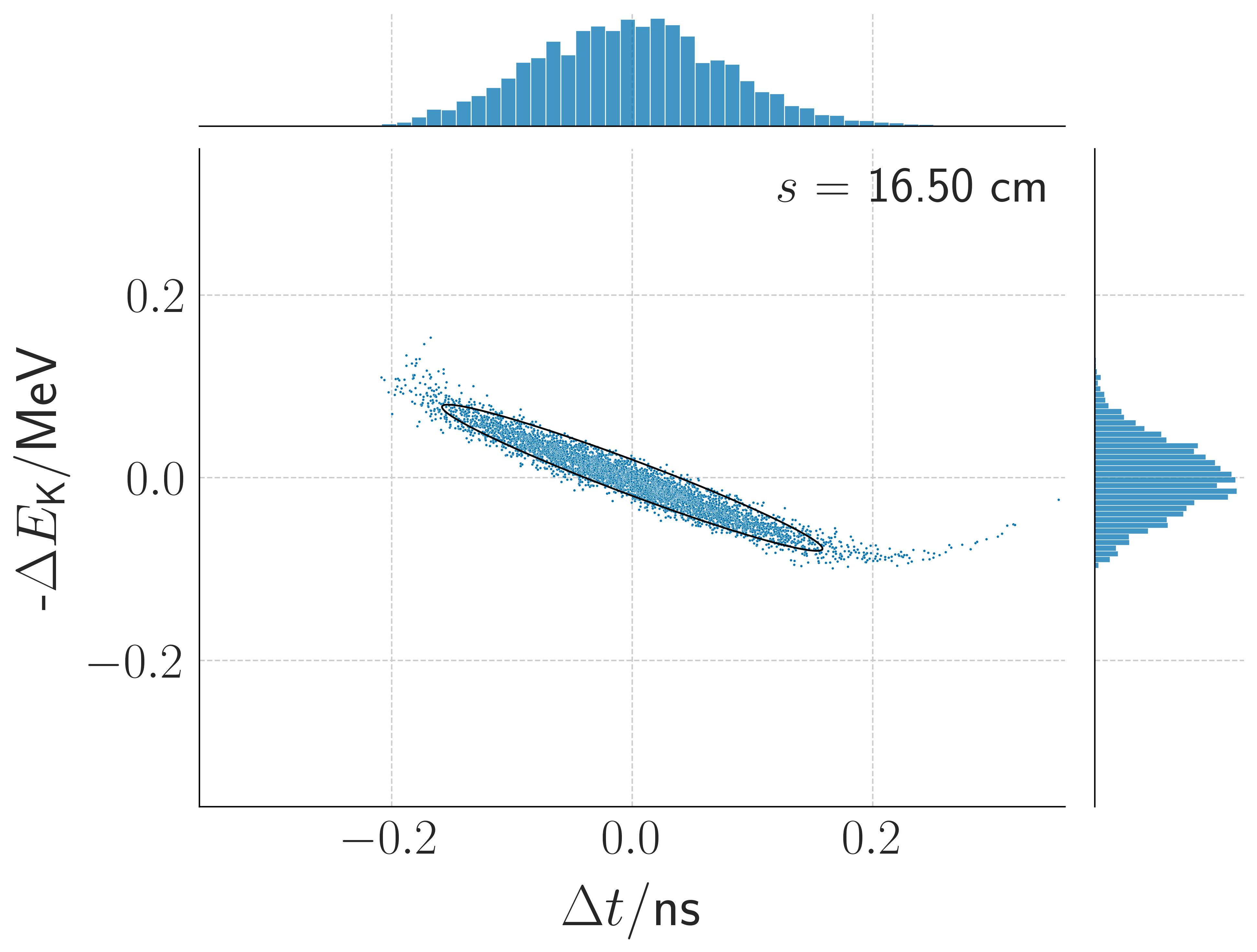}
     \caption{}
     \label{fig:ps-comparison-longitudinal-long}
    \end{subfigure}%
    \vspace*{\fill}
    \begin{subfigure}{0.49\textwidth}
     \centering
     \includegraphics[width=\textwidth]{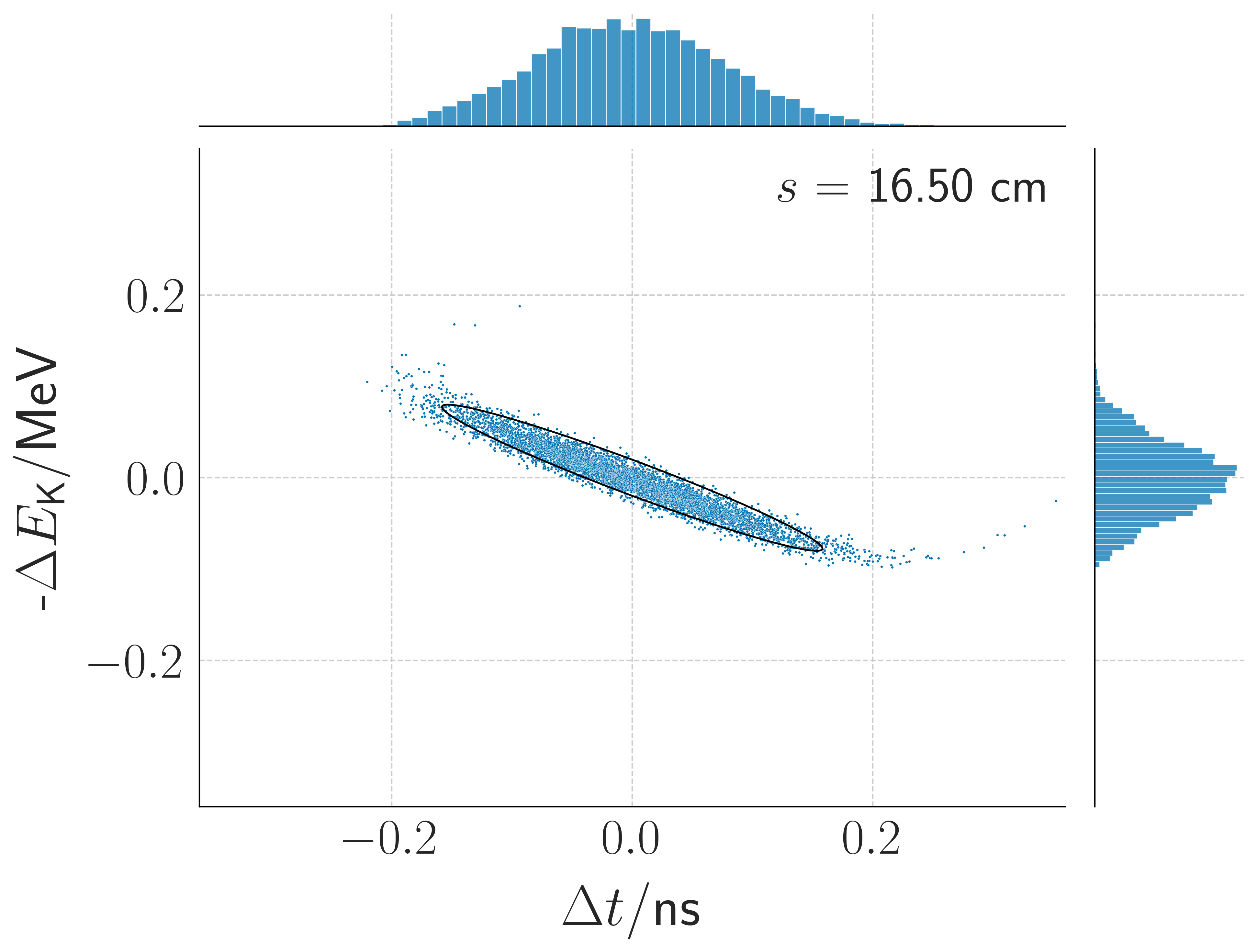}
     \caption{}
     \label{fig:ps-comparison-longitudinal-1um}
    \end{subfigure}%
    \hspace*{\fill}
    \begin{subfigure}{0.49\textwidth}
     \centering
     \includegraphics[width=\textwidth]{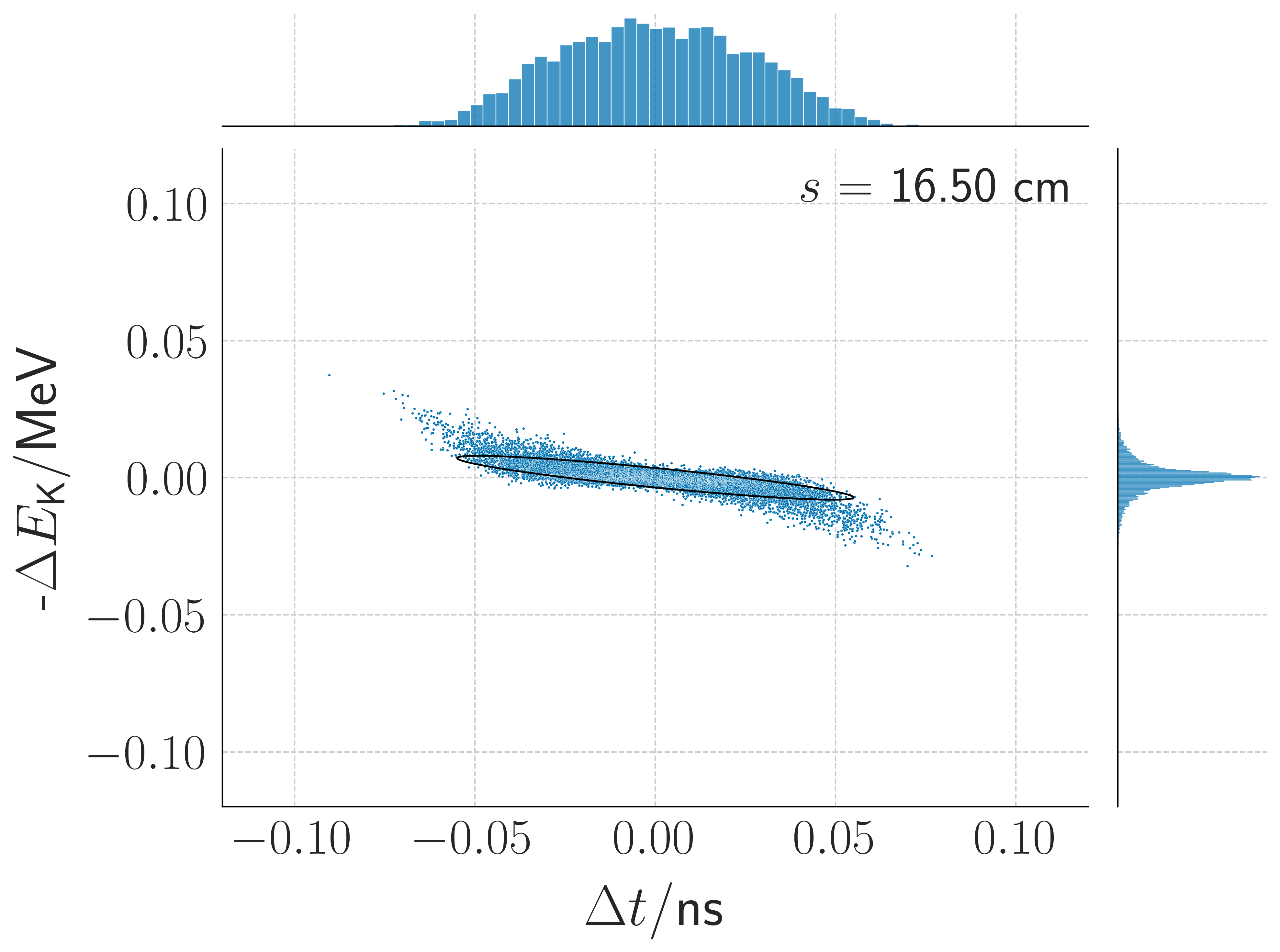}
     \caption{}
     \label{fig:ps-comparison-longitudinal-sc}
    \end{subfigure}%
    \caption{Longitudinal phase space at the exit of a 15~cm, 10~MV/m DWA segment. Particle data (blue) is from Warp and bounding ellipses (black) are reconstructed from \optr{} envelope data. Histograms above and to the right of each plot show the particle distributions in $\Delta t$ and $-\Delta E_\text{K}$, respectively. Each subplot shows the impact of altering a subset of parameters from the baseline rms normalized emittances of $\epsilon_{n,x} = \epsilon_{n,z} = 0.016$~$\mu$m and bunch charge $q_b = 1e$, while keeping the Courant-Snyder parameters constant. a) $\epsilon_{n,x} = 0.25$~$\mu$m, b) $\epsilon_{n,z} = 0.25$~$\mu$m, c) $\epsilon_{n,x} = \epsilon_{n,z} = 0.25$~$\mu$m, d) $q_b = 1\times10^8 e$.}
    \label{fig:ps-comparison-longitudinal}
\end{figure}

Increasing the transverse emittance up to $\epsilon_{n,x} = 0.25$~$\mu$m ($r = 0.25$~cm) did not degrade agreement between the \optr{} and Warp models, nor introduce observable nonlinear effects in either transverse (\cref{fig:ps-comparison-transverse-wide}) or longitudinal (\cref{fig:ps-comparison-longitudinal-wide}) phase space. Beyond this point, frequent wall interactions occurred ($\sim 1.5$\% of particles at $\epsilon_{n,x} = 0.5$~$\mu$m), precluding meaningful comparison with the linear model. 

In contrast, increasing the longitudinal emittance introduced deviations from linear behaviour in the Warp simulations. As shown in \cref{fig:ps-comparison-transverse-long}, the transverse phase space remains largely linear at $\epsilon_{n,z} = 0.25$~$\mu$m ($\Delta t = 0.125$~ns), but a subset of particles exhibit larger divergence values than expected, which is particularly noticeable at small $x$ values. In the longitudinal plane (\cref{fig:ps-comparison-longitudinal-long}), particles near the head and tail of the bunch exhibit reduced energy relative to the linear prediction. At the leading edge (extending beyond the \optr{} envelope toward $-\Delta t$), this produces enhanced phase slippage and a broadened energy spread. At the trailing edge (extending toward $+\Delta t$), it results in increased beam loss.

Simultaneous increases in both transverse and longitudinal emittance did not introduce qualitatively new behaviour. The transverse phase space (\cref{fig:ps-comparison-transverse-1um}) retains the geometric extent of the high-$\epsilon_{n,x}$ case, with the additional aberrations associated with large $\epsilon_{n,z}$. Similarly, the longitudinal phase space (\cref{fig:ps-comparison-longitudinal-1um}) is nearly identical to the high-$\epsilon_{n,z}$ case.

In the high-space charge case ($q_b = 10^8 \, e$), beam self-fields introduced cubic aberrations in both the transverse (\cref{fig:ps-comparison-transverse-sc}) and longitudinal (\cref{fig:ps-comparison-longitudinal-sc}) phase space planes. The overall bunch remained well behaved, with the core of the particle distribution evolving consistently with the linear prediction.

In the low energy source injection test, agreement between the two models was strong once the pulse duration was increased. However, it should be noted that the need for this adjustment was identified from the PIC simulations, where transit-time effects were apparent.

\begin{figure}
    \centering
    \begin{subfigure}{0.49\textwidth}
     \centering
     \includegraphics[width=\textwidth]{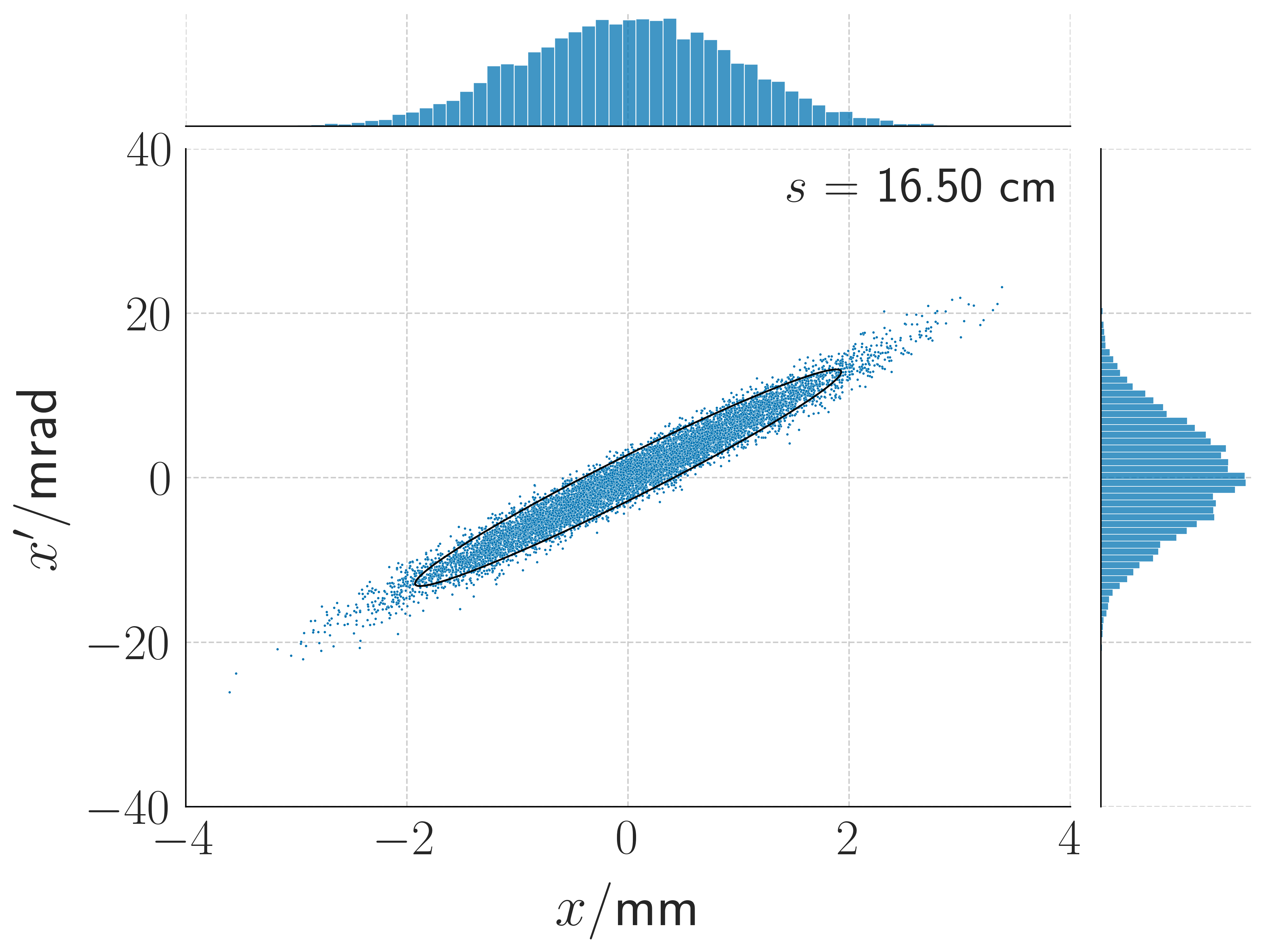}
     \caption{}
     \label{fig:ps-lowenergy-transverse}
    \end{subfigure}%
    \hspace*{\fill}
    \begin{subfigure}{0.49\textwidth}
     \centering
     \includegraphics[width=\textwidth]{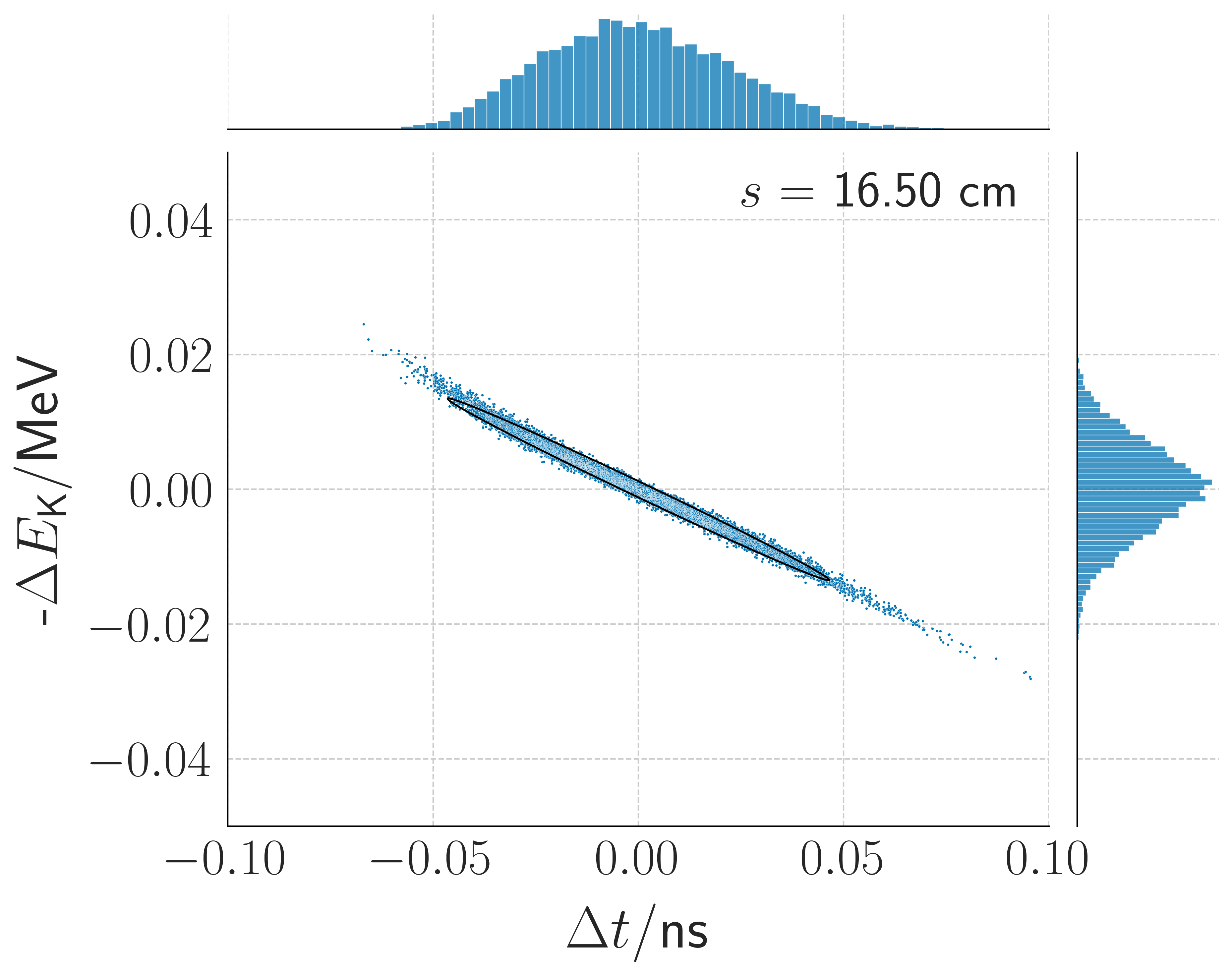}
     \caption{}
     \label{fig:ps-lowenergy-longitudinal}
    \end{subfigure}%
    \caption{Phase space distributions at the exit of a 15~cm, 10~MV/m DWA segment for a proton bunch injected at 30~keV. Particle data (blue) is from Warp and bounding ellipses (black) are reconstructed from \optr{} envelope data. Histograms above and to the right of each plot show the particle distributions along the corresponding axes. a) Transverse phase space, b) longitudinal phase space.}
    \label{fig:ps-lowenergy}
\end{figure}

\section{Discussion}\label{sec:discussion}

As in our previous work (\cite{lund-jung_2025}), the model presented here describes particle transport through prescribed axisymmetric fields in the DWA beam region. Although the electric and magnetic fields are now evaluated throughout the three-dimensional beam region, they do not yet include azimuthal structure or geometry-specific field perturbations. Similarly, the time profile does not yet reflect the output of realistic pulser circuits. In Warp, these fields are applied as external field elements, while the fields generated by the bunch are calculated when space charge is enabled. The applied accelerating fields do not respond dynamically to the beam or to the neighbouring acceleration modules. Consequently, effects such as beam-driven modification of the accelerating fields, cell-to-cell coupling, pulse timing errors and their interplay with nonlinear pulse shapes, and other practical effects are beyond the scope of this work. The beam loss, phase-space distortions, and other departures from linear transport identified here should therefore be interpreted as constraints arising within the prescribed field model, rather than as complete operating limits for a practical DWA system.

Nevertheless, we argue that this level of abstraction serves a useful role in the development of a new technology. The analytical field model provides a controlled basis for cross-checking the PIC implementation against the existing linear optics model before introducing additional hardware-specific effects. It can then serve as a basis for comparison for increasingly complex models. During prototype development, it allows beam-transport effects to be studied without requiring a complete accelerator geometry or pulsed-power design. This allows for exploratory beamline studies and the creation of idealized beam models for treatment-planning studies, both of which can provide valuable information for hardware design.

In the future, more detailed field descriptions could also be introduced as DWA designs and their electromagnetic models mature. The external-field implementation can, in principle, accommodate field maps generated using electromagnetic simulations or obtained from measurements. Empirical models of pulse deformation due to parasitic coupling effects could also be considered to modify the superposition of module fields. However, as previously noted, producing sufficiently complete and validated inputs will require further development beyond the scope of the present work.

Regarding the model itself, \cref{sec:theory} provides a natural extension of the analytical treatment introduced in our previous work (\cite{lund-jung_2025}). In particular, the leading term in the expansion of \cref{eq:evec-r} is directly related to Equation~11 in \lundjung, which was derived from the electric potential. The treatment of the magnetic field, however, represents a more substantial extension. The expression derived here is formally valid under the assumption of weak temporal variation; for strongly nonlinear time profiles, higher-order terms may become significant. Nonetheless, each successive term carries an additional factor of $1/c^2$, such that \cref{eq:bfield-crossproduct} is expected to dominate under realistic accelerating conditions.

From a computational standpoint, the evaluation of the magnetic field via the 3D integral in \cref{eq:bvec-theta} constitutes the most expensive component of the model. To linear order about the axis, the ratio of the magnetic contribution to transverse focusing relative to the electric component scales as $\beta$\tprime $\sfrac{\E_s}{\E_s'}$, where the prime denotes a derivative with respect to $s$. The quantity $\sfrac{\E_s}{\E_s'}$ is typically on the order of $10^{-2}-10^{-3}$~m. As a result, the magnetic contribution is generally small, and this calculation can often be neglected without significant loss of accuracy, particularly at low energies and for nearly flat temporal profiles. In practice, this approximation is most useful for short segments, modest accelerating gradients, and small values of \tprime{}, where excluding the magnetic field provides a substantial reduction in computational cost.

The spatial structure of the fields is consistent with physical expectations. The accelerating field $\mathcal{E}_s$ expands longitudinally as it propagates away from the wall, leading to a reduction in peak field strength while preserving the integral along $s$, as required by energy conservation in vacuum. The degree of field uniformity is governed by the beam pipe radius: smaller radii improve uniformity at the expense of tighter constraints on beam size. The radial field $\mathcal{E}_r$ exhibits the expected fringe-field behaviour of an open waveguide, providing focusing for $s < s_i$ and defocusing for $s > s_i$. The net effect on beam transport depends on the acceleration per module, which modifies both transit time and beam rigidity. For \tprime$ > 0$, this balance is partially offset by the increased field strength in the downstream (defocusing) region, although the superposition of fields from neighbouring modules complicates this picture~\cite{lund-jung_2025}.

The transport studies highlight several effects that warrant further investigation. Transverse beam size is strongly limited by the small aperture: initial 2-rms radii exceeding $0.25,r_w$ lead to frequent wall interactions, despite the presence of weak focusing fields. Given the likely fragility of thin dielectric beam pipes, such losses must be minimized in any practical design. Here we also see an example of the benefit of particle-resolved simulations. While the \optr{} envelope calculation can indicate the beam size as it approaches an aperture and allow for an inference on beam loss, the particle-resolved Warp simulations directly identify the fraction and phase-space location of particles lost at the wall.

Longitudinally, increasing bunch length introduces a characteristic distortion of the phase space distribution, with reduced particle energy at both the head and tail of the bunch. At the leading edge, this produces enhanced phase slippage and increased energy spread, while at the trailing edge it leads to particle loss as the particles lose coordination with the virtual travelling wave. These effects are reminiscent of ``ear fields'' observed in induction accelerators~\cite{takayama2011induction}, as well as the longitudinal stability limits encountered in RF accelerators. Mitigation strategies include increasing the accelerating gradient or reducing the bunch length, both of which improve longitudinal confinement. In the context of DWAs, an additional degree of freedom exists through the adjustment of \tprime{}, although this introduces trade-offs with transverse dynamics. Furthermore, the form of \tprime{} considered in this work is already idealized, and may be difficult to achieve in practice. These data indicate that maintaining longitudinal stability, especially in cases where particles extend towards field edges, may require particular attention in practical designs.

For the charge and emittance values considered here, the linear model provides an accurate and computationally efficient description of beam transport. However, the PIC simulations reveal more stringent constraints on longitudinal beam quality and pulse duration than would be inferred from the linear model alone. These findings have important implications for DWA-based accelerator design. In contrast to conventional systems, where beam losses may be tolerated and managed through collimation or bending sections, DWA concepts are often imagined as straight-through systems delivering beam directly to the patient. Under such conditions, any sources of beam loss and contamination from wall collisions must be minimized. This places strong requirements on injection conditions, particularly in terms of bunch length and transverse size. While downstream correction or filtering systems may be conceivable, they cannot be assumed in early-stage designs. The continued development of particle-resolved models or the development of higher-order transfer map models of DWA transport may be important in identifying these limits in any practical design.

More broadly, the results highlight the complementary roles of linear optics and PIC modelling in DWA development. Linear models enable rapid exploration of design space and parameter optimization, while PIC simulations provide additional information on particle loss, phase space distortions, and higher order effects that can be important in identifying beam-dynamics constraints. Together, these tools provide a pathway toward studying the beam-dynamics requirements of practical DWA-based systems, where beam quality, stability, and robustness must be balanced against the constraints of compact, high-gradient acceleration.

\section{Conclusion}\label{sec:conclusion}

In this work, an analytical field model was developed for the study of DWA beam dynamics. Starting from a prescribed on-wall excitation, the three-dimensional, time-dependent axisymmetric electric and magnetic fields of arbitrary DWA modules were derived and implemented in the open-source Python package pyDWA. The model was integrated with the particle-in-cell code Warp by implementing the modules as external field elements. This extends our previous linear optics treatment to a particle-resolved calculation while retaining the flexibility of an analytical field description. In the absence of experimental beam-transport data, the implementation was cross-checked against our linear optics treatment under deliberately controlled conditions where transport is expected to remain predominantly linear. Close agreement was observed between the two models in this regime.

The simulations were then extended to less idealized beam conditions. In each case, the particle distributions from Warp were examined directly and compared with the corresponding \optr{} envelope predictions. Increasing the transverse emittance increased the likelihood of beam loss through wall interactions, but overall envelope evolution remained consistent between the models. Increasing the longitudinal emittance produced phase-space distortions and particle losses due to longitudinal instability. Again however, envelope evolution remained consistent over the parameter range studied. At 1$\times$10$^8e$, the largest bunch charge considered, the bunch remained well behaved, although the beam self-fields introduced a visible cubic aberration in both transverse planes. Simulations at a lower injection energy showed good agreement between Warp and \optr{}.

These results support the use of the linear optics model as a fast tool for studying DWA beam transport. They also suggest the usefulness of the PIC implementation. By resolving the particle distribution directly, the PIC model makes it possible to directly identify losses, phase-space distortions, and higher-order effects. This additional layer of detail provides an important complement to linear optics studies. Finally, it should be noted that the constraints identified here are specific to transport through the prescribed axisymmetric fields, and should not be interpreted as operating limits for a practical DWA system. Future work will consider beam preparation, pulse timing and other field errors, cell-to-cell coupling, and increasingly detailed field descriptions as suitable experimental or simulation data become available.

\section{Acknowledgements}
This work was supported by the Fonds de recherche du Qu\'{e}bec – Nature et technologies and by a project grant from the Canadian Institutes of Health Research (grant \# CIHR-PJT 183753). TRIUMF is funded under a contribution agreement with the National Research Council Canada.

\section{Conflict of Interest Statement}
The authors have no relevant conflicts of interest to disclose.

\newpage

\bibliography{paper}

\end{document}